\documentclass[10pt,a4paper,twocolumn]{IEEEtran}
\usepackage{cite}
\usepackage{amsmath,amssymb,amsfonts}
\usepackage{algorithmic}
\usepackage{graphicx}
\usepackage{textcomp}

\def\BibTeX{{\rm B\kern-.05em{\sc i\kern-.025em b}\kern-.08em
    T\kern-.1667em\lower.7ex\hbox{E}\kern-.125emX}}
\begin{document}
\setlength{\topsep}{0.0cm}
\setlength{\itemsep}{0.0em}
\title{{\large Identification of LFT Structured Descriptor Systems with Slow and Non-uniform Sampling}}
\vspace{-0.5cm}
\author{{\small Tong Zhou, \IEEEmembership{Fellow, IEEE}}
\thanks{ This work was supported in part by the NNSFC, China under Grant
	 62373212, 62127809, 61733008, and 52061635102, and by the BNRist project under Grant BNR2024TD03003. }
\thanks{T. Zhou is with
the Department of Automation and BNRist, Tsinghua University, Beijing, 100084, China (e-mail: tzhou@mail.tsinghua.edu.cn).}
\vspace*{-1.0cm}}

\maketitle
\vspace{-0.5cm}
\begin{abstract}
Time-domain identification is studied in this paper for parameters of a continuous-time multi-input multi-output descriptor system, with these parameters affecting system matrices through a linear fractional transformation. Sampling is permitted to be non-uniform and slow, in the sense that it is not required to satisfy the Nyquist frequency restrictions. This model can be used to describe the behaviors of a networked dynamic system, and the obtained results can be straightforwardly applied to a standard state-space model, as well as a lumped system. An explicit formula is obtained respectively for the transient and steady-state responses of the system stimulated by a signal generated by another system. Some relations have been derived between the system steady-state response and its transfer function matrix (TFM), which reveal that the value of a TFM at almost any interested point, as well as its derivatives and a right tangential interpolation along an arbitrary  direction, can in principle be estimated from time-domain input-output experimental data. Based on these relations, an estimation algorithm is suggested respectively for the parameters of the descriptor system and the values of its TFM. Their properties like asymptotic unbiasedness, consistency, etc., are analyzed. A simple numerical example is included to illustrate characteristics of the suggested estimation algorithms.
\end{abstract}


\addtolength{\abovedisplayskip}{-0.15cm}
\addtolength{\belowdisplayskip}{-0.15cm}

\addtolength{\abovedisplayshortskip}{-0.15cm}
\addtolength{\belowdisplayshortskip}{-0.15cm}

\setlength{\footskip}{-1.0cm}
\addtolength{\footskip}{-1.40cm}

\addtolength{\topmargin}{-0.10cm}
\addtolength{\textheight}{0.62cm}

\vspace*{-0.5cm}
\section{Introduction}
\label{sec:introduction}

\setlength{\itemsep}{0.0em}
\IEEEPARstart{I}{n} describing the dynamics of a networked dynamic system (NDS), descriptor systems have been proved to be a powerful model. This model has been extensively adopted in various areas like electric engineering, chemical engineering processes, social systems, ecosystems, and so on \cite{abg2020,Dai1989,Duan2010,hv2004,ptt2019}. Typical examples include constrained mechanical systems \cite{Duan2010}, electrical power grids \cite{ptt2019,abg2020}, the dynamic Leontief model for an economic system \cite{Dai1989}, oil catalytic cracking process \cite{abg2020,Dai1989}, etc. An attractive property of this model is its capabilities of describing relations among different natural variables, while allowing  an explicit expression for system working principles. Both transfer function matrix (TFM) and descriptor form state-space model are widely used in describing these system's dynamics. When a large-scale complex descriptor system is under investigation, it is usually constituted from numerous subsystems, and its behaviors are determined not only by subsystem dynamics, but also by their  interactions, while the latter is usually called the structure of an NDS \cite{cw2020,Zhou2022,zy2022}. Revealing the dynamics and structure of an NDS from experimental data has been attracting extensive attentions from various field, such as data analysis and processing, system analysis and design, etc. Recently, system scale increases rapidly due to developments in computer science, communication technologies and artificial intelligence, etc., which greatly stimulates researches in NDS structure estimations \cite{cw2020,dtrf2019,Fortunato2010,sp2015,wvd2018,ylwv2020,Zhou2022,zg2012}.

Concerning NDS structure, a state space model is argued in \cite{cw2020} to be more informative than a TFM model. While a state space model can be derived in many applications using first principles, cases also exist in which some critical parameters are difficult, or even impossible, to be directly measured. Examples include protein activities in a biological system, substrate concentrations of a chemical reaction process, etc. The problem of estimating these parameters with measurements then arises, which plays a vital role in revealing NDS dynamics and structure that is also closely related to machine learning interpretability \cite{vtc2021,ptt2019,Ljung1999,wvd2018,zly2024,zy2022}.

In \cite{Zhou2022,zy2022}, system matrices of an NDS are shown in general to be a linear fractional transformation (LFT) of its subsystem connection matrix (SCM), while the latter describes subsystem interactions. In addition, when system matrices of a subsystem depend on its  parameters through an LFT, all these parameters can also be included in a modified SCM through introducing some virtual internal inputs and outputs. These conclusions clarify that an LFT structured descriptor system can include a wide class of NDSs as a special case, and system matrix elements or TFM coefficients of an NDS are generally not algebraically independent.

Until now, most researches in system identification are focused on black-box discrete-time system with uniform sampling \cite{Ljung1999,ps2001}. In actual applications, there are various situations in which system working principles must be taken into account in building its model, uniform sampling is quite expensive or even impossible, the Nyquist frequency restriction can hardly be satisfied in sampling, and continuous-time model is greatly anticipated \cite{cw2020,pglj2019,ylwv2020,ytlyg2021,zdg1996,zly2024}. Only a few results, however, have been reported for these cases. Under the assumption that all the system matrices depend affinely on its parameters, \cite{mpr2014} clarifies mathematical difficulties in estimating parameters for a structured state-space model, and suggests some procedures for selecting the initial value in a nonlinear optimization associated with this estimation. \cite{ylwv2020} investigates identification of a discrete-time state-space model with its system matrices depending linearly or polynomially on its parameters. Ideas are borrowed from subspace identifications and nuclear norm optimizations, in which the associated non-convex optimization problem is converted into a sequential convex programming. In \cite{wvd2018}, some methods are suggested to testify identifiability of an NDS, regarding each state of the NDS as a node, while influences among its states as edges. \cite{vtc2021} studies how to recover subsystem interactions from the estimated impulse responses of a specific kind of discrete-time NDSs, and some identification procedures have been developed. \cite{ytlyg2021} clarifies that a low sampling period in general leads to a discrete-time model with a structure significantly different from its original continuous-time counterpart.  \cite{Zhou2022} shows that an NDS is always identifiable under the constraint that the TFM from its internal inputs to its external outputs is of full normal column rank (FNCR), while the TFM from its external inputs to its internal outputs is of full normal row rank (FNRR). \cite{zly2024} investigates identifiability for descriptor systems with an LFT structure in the frequency domain, and derives a condition that can be verified recursively with each prescribed frequency.

Recently, a so-called invariant-subspace based identification method is suggested in \cite{hfhw2022} to deal with continuous-time system identification with slow and non-uniform sampling. Based on some results in output regulation theories for a nonlinear system, an estimation procedure is suggested for a black-box transfer function model of a single-input single-output (SISO) linear time invariant (LTI) continuous-time system, under the situation that the probing signal is restricted to be a multi-sine function. These results are further extended by the authors to a multi-input multi-output (MIMO) state-space model, and a nonlinear dynamic system with polynomial nonlinearities, without either alleviating restrictions on the probing signals, or taking system working principles into account.

In systems and control theories, operator theories, etc., interpolation conditions are extensively utilized in settling various important theoretical issues, such as constructing a rational function, estimating a system model, and reducing model complexities, etc. \cite{abg2020,bgr1990,bgd2020,msa2024}. It is still, however, not very clear how to obtain these conditions from experimental data. As a byproduct of this study, a relation has been established between steady-state response of a system and its right interpolations, which makes their data-driven estimations in principle possible.

In this paper, we investigate parameter estimation for a continuous-time LTI MIMO descriptor system with an LFT structure, directly using measured time-domain system output samples. This model includes all the aforementioned LTI models as a special case. The sampling is required neither to be uniform nor faster than the Nyquist frequency, and the probing signal can be an arbitrary time function generated by an LTI system. Explicit formulas have been respectively derived for the system transient and steady-state responses. It has been made clear that steady-state system response only is able to provide complete information of the value of the system TFM at some finitely many  points of the complex plane. In addition, it is in principle possible to simultaneously estimate the value of the system TFM at almost any points, its derivatives with respect to the Laplace variable, as well as its right tangential interpolations along any directions. Based on these observations, a least squares based estimate is at first derived for simultaneously identifying TFM values of the descriptor system at several particular real/complex points, from which an estimate for the system parameter vector is further obtained. Their convergence properties have also been established, as well as condition for persistent excitations. These results are directly applicable to a standard state-space model in which system matrices depend on its parameters through an LFT. To illustrate advantages of the obtained theoretical results and the suggested estimation algorithms, a simple mass-spring-damper example is included, in which parametric estimation accuracies are compared with the ordinary least squares based estimate. It has been observed that the well known local minimum difficulty occurs also in this simple example with a direct nonlinear data-fitting, and is successfully avoided by the suggested estimation procedure.

The remaining of this paper is organized as follows. Section \ref{sec:problem} describes the model used in this paper and gives a precise statement of the problem. Section \ref{sec:res-dec} investigates how to explicitly decompose a system response into its transient and steady-state parts, clarifying individual influences respectively from system initial conditions and input signals. Estimation procedures are given in Section \ref{sec:par-est}, while its properties are investigated in Section \ref{sec:par-pro}, together with requirements from data informativity. Section \ref{sec:example} reports some numerical simulation results. Finally, some concluding remarks are provided in Section \ref{sec:conclusions}. An appendix is included to give proofs of some technical results, while another appendix to discuss system output decompositions for a particular case, in which the transition matrix of the input generating system (IGS) has only one real Jordan block.

The following notation and symbols are adopted in this paper. $ \mathbb{R}^{n}/\mathbb{C}^{n} $ and $ \mathbb{R}^{m\times n} $ represent respectively the sets of $ n $ dimensional real/complex vectors and $ m\times n $ dimensional real matrices, while $||\cdot||$ the Euclidean norm of a vector or its induced norm of a matrix. The superscript is usually omitted when it is equal to 1. $ \star^{\perp}_{r} $ and $ \star^{\perp}_{l} $ stand respectively for the matrix constituted from a basis of the right and left null spaces of a matrix, while $\overline{\sigma}(\star)$ and $\underline{\sigma}(\star)$ respectively the maximum and minimum of its nonzero singular values. $ {\it tr}\!\left\lbrace \star \right\rbrace$ denotes the trace of a matrix, while $ \star^{T} $ and $ \star^{\dagger} $ respectively its transpose and Moore-Penrose pseudo inverse. $ {\it vec}\!\left\lbrace \star \right\rbrace  $ represents vectorization of a matrix, and ${\it diag}\!\{\star_{i}\:|^n_{i=1}\}$, $ {\it  col}\!\left\lbrace \star_{i}\:|_{i=1}^{n} \right\rbrace  $ and $ {\it row}\!\left\lbrace \star_{i}\:|_{i=1}^{n} \right\rbrace $ the matrices composed of $\star _{i}|_{i=1}^{n}$ stacking respectively diagonally, vertically and horizontally. For a complex variable/vector/matrix, the superscript $*$, $[r]$ and $[i]$ denote respectively its conjugate, real part and imaginary part, while $ {\it R_{e}}\left\lbrace\cdot\right\rbrace$ and $ {\it I_{m}}\left\lbrace\cdot\right\rbrace$ respectively the operations of taking its real and imaginary parts. $I_{n} $ stands for the $ n $ dimensional identity matrix, while $ 0_{m\times n} $ the $ m\times n $ dimensional zero matrix. When the dimensions are obvious or insignificant, their subscripts are usually omitted. For a random variable/vector/matrix, ${\it E_{x}}\{\cdot\}$ and $ {\it P_{r}}\left\lbrace\cdot\right\rbrace$ stand respectively for the operations of taking its mathematical expectation and calculating the probability of an associated event. ${\cal L}(\cdot)$ is adopted to denote the Laplace transformation of a vector valued function of time, while ${\cal L}^{-1}(\cdot)$ its inverse transformation.

\vspace{-0.25cm}
\section{Problem Formulation and Preliminaries}\label{sec:problem}

Consider an LTI continuous-time descriptor system $\mathbf{\Sigma}_{p}$ that has $m_{\theta}$ parameters $\theta_{i} | _{i=1}^{m_{\theta}}$ to be estimated. Define a vector $\theta$ as $\theta = col \left\lbrace \theta_{i} | _{i=1}^{m_{\theta}} \right\rbrace $, and assume that the input-output relations of the descriptor system $\mathbf{\Sigma}_{p}$ can be described as
\begin{eqnarray}
	E \dot{x}(t)  &=& A(\theta)x(t)+B(\theta)u(t)\label{plant-1}\\
	y(t) &=& C(\theta)x(t)+D(\theta)u(t)   \label{plant-2}
\end{eqnarray}
\hspace*{-0.00cm}Here, $x(t)\in\mathbb{R}^{m_{x}}$, $u(t)\in\mathbb{R}^{m_{u}}$ and $y(t)\in\mathbb{R}^{m_{y}}$ are respectively the system state, input and output vectors. Moreover, assume that the system matrices $A(\theta)$, $B(\theta)$, $C(\theta)$ and $D(\theta)$ depend on its parameters through the following LFT form,
\begin{eqnarray}\label{plant-3}
		\left[\!\!\!\begin{array}{cc}
				A(\theta) & B(\theta)\\
				C(\theta) & D(\theta)
		\end{array}\!\!\!\right] \!\!\!\!\!\!
&=& \!\!\!\!\!\!
		\left[\!\!\!\begin{array}{cc}A_{xx} & B_{xu}\\
				C_{yx} & D_{yu}\\
		\end{array}\!\!\!\right] \!\!+\!\!
        \left[\!\!\begin{array}{c}
				B_{xv}\\
				D_{yv}
        \end{array}\!\!\right]\!
        \left[ I_{m_{v}} \!\!-\!\! P(\theta)D_{zv}\right]^{-\!1}\!\!\times \nonumber\\
		& &\!\!\!\!\!\!  P(\theta)
        \left[\begin{array}{cc}
				C_{zx} & D_{zu}
        \end{array}\right]
\end{eqnarray}
\hspace*{-0.00cm}in which
\begin{equation}\label{plant-4}
	P(\theta)=\sum\limits_{i=1}^{m_{\theta}}\theta_{i}P_{i}
\end{equation}
\hspace*{-0.00cm}In addition, assume that a set $\mathbf{\Theta}$ is given that consists of all possible parameter values determined by some a priori plant information. In this description, it is also assumed that $E\in \mathbb{R}^{m_{x}\times m_{x}}$, $A_{xx}\in \mathbb{R}^{m_{x}\times m_{x}},\; B_{xu}\in \mathbb{R}^{m_{x}\times m_{u}},\; B_{xv}\in \mathbb{R}^{m_{x}\times m_{v}},\; C_{yx}\in \mathbb{R}^{m_{y}\times m_{x}},\; C_{zx}\in \mathbb{R}^{m_{z}\times m_{x}},\; D_{zu}\in \mathbb{R}^{m_{z}\times m_{u}},\; D_{zv}\in \mathbb{R}^{m_{z}\times m_{v}},\; D_{yu}\in \mathbb{R}^{m_{y}\times m_{u}}$ and $D_{yv}\in \mathbb{R}^{m_{y}\times m_{v}}$, as well as  $P_{i}\in \mathbb{R}^{m_{v}\times m_{z}}$ with $i=1, 2, \cdots, m_{\theta}$, are some prescribed matrices, reflecting available a priori knowledge about the descriptor system from its working principles, etc.

On the other hand, assume that the input signal $u(t)$ of the above descriptor system is generated by the following autonomous LTI system $\mathbf{\Sigma}_{s}$, with its state vector $\xi(t)$ belonging to $\mathbb{R}^{m_{\xi}}$ and its system matrices $\Xi$ and $\Pi$ having compatible dimensions, that is $\Xi \in \mathbb{R}^{m_{\xi}\times m_{\xi}}$ and $\Pi \in \mathbb{R}^{m_{y}\times m_{\xi}}$,
\begin{equation}
	\dot{\xi}(t)  = \Xi \xi(t), \hspace{0.5cm}
	u(t) = \Pi \xi(t)  \label{signal-1}
\end{equation}

The objectives of this paper are to develop an estimation procedure for the parameter vector $\theta$, using measured outputs of the descriptor system $\mathbf{\Sigma}_{p}$ at some sampling instants, denote them by $y_{m}(t_{k})|_{k=1}^{N}$, under the condition that all the system matrices of the IGS $\mathbf{\Sigma}_{s}$ and its initial condition vector $\xi(0)$, are exactly known.

It is worthwhile to emphasize that in the above problem, the sampling instants are not required to be uniformly distributed, while the sampling rate is not asked to be faster than the Nyquist frequency. In addition, all the obtained results can be directly applied to a system with its input-output relations being described by a standard state-space model. That is, the model of Equations (\ref{plant-1}) and (\ref{plant-2}) with $E=I_{m_{x}}$.

In the above system description, the parameters to be estimated, i.e. $ \theta_{i}, i=1,\cdots,m_{\theta} $, can in general be some unknown values of its first principle parameters (FPP), such as those of resistors, capacitors and inductances in a circuit, temperatures and substance concentrations in a chemical process, etc. They can also be some functions of these FPPs that hold some application significance in system analysis and synthesis, such as time constants and steady-state gains of a dynamic system, etc. To make identification of these parameters meaningful, it is essential to assume that they are algebraically independent of each other, meaning that they do not satisfy any non-trivial polynomial equation with real coefficients \cite{Dai1989,Duan2010,Zhou2022,zly2024}.

It is now well known that system matrices of a large class of LTI systems depend on their FPPs through an LFT \cite{Ljung1999,cw2020,zdg1996}. In addition, \cite{Zhou2022} and \cite{zy2022} make it clear that through introducing some virtual internal inputs and outputs in a subsystem, system matrices of an NDS can in general be expressed as an LFT of its subsystem parameters and interactions. These mean that while the adopted assumptions on the descriptor system $\mathbf{\Sigma}_{p}$ do restrict its applicability, the associated results may still be meaningful in various application significant cases.

In this paper, the matrix $E$ of the descriptor system $\mathbf{\Sigma}_{p}$ is assumed known and independent of its parameters $\theta_{i}|_{i=1}^{m_{\theta}}$, which is different from its other system matrices $A(\theta)$, $B(\theta)$, $C(\theta)$ and $D(\theta)$, whose elements are functions of these parameters. Noting that this matrix is usually adopted to describe constraints among natural variables in applications \cite{Dai1989,Duan2010}, this assumption does not appear very restrictive. However, it is worthwhile to emphasize that there do exist cases in which the matrix $E$ depends on some system parameters \cite{Dai1989,Duan2010,hv2004}. Nevertheless, \cite{hv2004} has clarified that even in this case, through some fairly straightforward algebraic manipulations, the TFM of the descriptor system $\mathbf{\Sigma}_{p}$ can still be expressed as an LFT of its parameters under some quite weak conditions, which has completely the same form as that of the following Equation (\ref{decom-5}). These mean that by the same token developed in this paper, similar parameter estimation algorithms can be developed for the descriptor system $\mathbf{\Sigma}_{p}$, even when its system matrix $E$ also varies with its parameters. To avoid awkward arguments, this extension is not attacked in this paper.

To clarify dependence of the descriptor system $\mathbf{\Sigma}_{p}$ on its parameters, it is sometimes also written as $\mathbf{\Sigma}_{p}(\theta)$. Similar expressions have also been adopted for other symbols, such as its output vector $y(t)$, its TFM $H(s)$, etc.

To make parameter estimations meaningful for the aforementioned descriptor system, it is necessary that its output vector does not depend on its future input vector, and each pair of admissible initial conditions and admissible excitation signals can generate one and only one output signal. This means that the following two assumptions are necessary for investigating this identification problem.

\newtheorem{Assumption}{Assumption}

\begin{Assumption}\label{assum:1}
For each $\theta\in\mathbf{\Theta}$, the descriptor system $\mathbf{\Sigma}_{p}$ is regular, meaning that the matrix valued polynomial $s E-A(\theta)$ is invertible. Here, $s$ stands for the Laplace transform variable.
\end{Assumption}

\begin{Assumption}\label{assum:2}
For each $\theta \!\!\in\!\! \mathbf{\Theta}$, the descriptor system $\mathbf{\Sigma}_{p}$ is well-posed, meaning invertibility of the matrix $I_{m_{v}} \!\!-\!\! P(\theta)D_{zv}$.
\end{Assumption}

In addition to these two assumptions which are widely adopted in descriptor system analysis and synthesis \cite{abg2020,Dai1989,Duan2010}, the following three assumptions are also adopted in this paper for developing a parameter estimation algorithm.

\begin{Assumption}\label{assum:3}
The descriptor system $\mathbf{\Sigma}_{p}$ is stable, and an upper bound is available for its settling time, denote it by $\overline{t}_{s}$.
\end{Assumption}

\begin{Assumption}\label{assum:4}
All the eigenvalues of the state transition matrix $\Xi$ of the IGS $\mathbf{\Sigma}_{s}$ are distinct from each other,  and have a real part not smaller than zero.
\end{Assumption}

\begin{Assumption}\label{assum:5}
Each sampling instant $t_{k}$ is not smaller than $\overline{t}_{s}$, $k=1,2,\cdots,N$.  The composite influences of process disturbances, measurement errors, etc., on the system output vector $y(t)$, can be described by a random sequence $n(t_{k})$. That is, $y_{m}(t_{k}) = y(t_{k}) + n(t_{k})$. Moreover, this random sequence is uncorrelated between any two sampling instants. Furthermore, ${\it E_{x}}\{n(t_{k})\}=0$ and ${\it E_{x}}\{n(t_{k})n^{T}(t_{k})\} \leq {\Sigma}_{n}$, with ${\Sigma}_{n}$ being a constant matrix having a finite maximum singular value.
\end{Assumption}

While Assumption \ref{assum:3} is necessary for performing an identification experiment in open loop, Assumptions \ref{assum:4} and \ref{assum:5} are adopted only for avoiding a complicated presentation, that may hide the main ideas behind the suggested estimation procedures. From the following derivations, it is not too difficult to understand that results of this paper can be straightforwardly extended to cases when either of these two assumptions, or both of them are not satisfied.

The following concept is a direct extension of frequency-domain identifiability of \cite{Ljung1999,ps2001,zly2024}, and plays important roles in analyzing properties of the suggested estimates.

\newtheorem{Definition}{Definition}

\begin{Definition}\label{def:1}
The descriptor system $\mathbf{\Sigma}_{p}$ is called identifiable with its TFM values at $s_{k}|_{k=1}^{m}$ with $s_{k} \in \mathbb{C}$, if the value of its parameter vector can be uniquely determined by them.
\end{Definition}

It becomes clear now that parameter identifiability is in general a generic property, meaning that rather than a particular value of the parameter vector, it is the ways that these parameters change the system's TFM that determine whether they are identifiable \cite{Ljung1999,ps2001,Zhou2022,zly2024}. This property makes it possible to develop a graph based procedure for identifiability verification \cite{Ljung1999,zy2022}.

The next well-known results in matrix analysis and computations are helpful in investigating persistent excitation conditions of the identification problem \cite{gv1989,zly2024}.

\newtheorem{Lemma}{Lemma}

\begin{Lemma}
Partition a matrix $A$ into $A \!\!=\!\! \left[ A_{1}^{T} \;  A_{2}^{T}\right]^{T}\!$. Assume that the submatrix  $A_{1}$ is not of full column rank (FCR). Then the matrix $A$ is of FCR, if and only if the matrix $A_{2}A_{1,r}^{\perp}$ is.
\label{lemma:1}
\end{Lemma}

The following results are extensively applied to convergence analysis of an estimation procedure, in which the first equation is widely known as the Chebyshev's inequality, while the second as the Borel-Cantelli's lemma in probability theories \cite{ct1997,Ljung1999,ps2001}.

\begin{Lemma}
Let $\zeta$ be a real valued random variable, and $\zeta_{k}$ with $k=1,2,\cdots$, be a real valued random sequence. Then for an arbitrary positive $\varepsilon$, the next inequality is valid.
\begin{equation}\label{eqn:lemma:0-1}
{\it P_{r}}\left\lbrace |\zeta| > \varepsilon \right\rbrace \leq \frac{{\it E_{x}}\left\lbrace \zeta^{2} \right\rbrace}{\varepsilon^{2}} \end{equation}
Moreover, if for every $\varepsilon \!>\! 0$, the next inequality is satisfied,
\begin{equation}\label{eqn:lemma:0-2}
 \sum_{k=1}^{\infty}{\it P_{r}}\left\lbrace |\zeta_{k}| > \varepsilon \right\rbrace < \infty 	
\end{equation}
Then the random sequence $\zeta_{k}$ converges to $0$ with probability 1 (w.p.1) with the increment of the index $k$.
\label{lemma:0}
\end{Lemma}

\vspace{-0.25cm}
\section{System output decomposition}\label{sec:res-dec}

To develop an estimation procedure for the parameters of the descriptor system $ \mathbf{\Sigma}_{p} $, a decomposition is derived for its output vector $y(t)$ in this section, which clarifies some attractive relations between its steady-state responses and its TFM. To simplify expressions, dependence of its system matrices $A(\theta)$, $B(\theta)$, $C(\theta)$ and $D(\theta)$ on its parameter vector $\theta$ is omitted throughout this section, as well as its TFM $H(s,\theta)$.

For this purpose, the following results are at first established, while its proof is deferred to Appendix I.

\newtheorem{Theorem}{Theorem}
\begin{Lemma}\label{lemma:2}
Assume that the descriptor system $ \mathbf{\Sigma}_{p} $ satisfies Assumptions \ref{assum:1} and \ref{assum:2}. Let $x(0)$ denote its initial state vector, while $\xi(0)$ that of the IGS $ \mathbf{\Sigma}_{s}$. Then for arbitrary $x(0)$ and $\xi(0)$, there exist a constant vector $\overline{x}(0)$ and a constant matrix $X$, such that at each time instant $t \geq 0$, the state vector $x(t)$ of the descriptor system $ \mathbf{\Sigma}_{p} $ can be expressed as
\begin{equation}\label{decom-1-x}
	x(t) = {\cal L}^{-1}\!\!\left\{(sE \!-\! A)^{-1}\right\}\overline{x}(0) \!+\! X \xi(t) 	
\end{equation}
if and only if there is a constant matrix $Z$, such that the following two matrix equations are simultaneously satisfied,
\begin{equation}\label{decom-1}
	EX - Z = 0, \hspace{0.5cm} AX + B\Pi - Z\Xi = 0 	
\end{equation}
In addition, when these requirements are satisfied, the vector $\overline{x}(0)$ is equal to $Ex(0) \!-\! Z\xi(0)$.
\end{Lemma}

It is interesting to note that when Equation (\ref{decom-1}) has a solution, straightforward algebraic manipulations show that \begin{displaymath}
E \frac{d[x(t)-X\xi(t)]}{dt} = E\dot{x}(t) - Z\dot{\xi}(t) = A[x(t)-X\xi(t)]
\end{displaymath}
That is, the time-dependent vector-valued-function $x(t) \!-\! X\xi(t)$ can be regarded as the state vector of an autonomous descriptor system. From this relation, Equation (\ref{decom-1-x}) can be directly established.

From Lemma \ref{lemma:2} and Equation (\ref{plant-2}), it is clear that when the conditions of Equation (\ref{decom-1}) are satisfied,
the output vector $y(t)$ of the descriptor system $ \mathbf{\Sigma}_{p} $ can be expressed as
\begin{eqnarray}
& & \label{decom-2}
	y(t) = y_{t}(t) + y_{s}(t)  \\
& & \label{decom-2-1}
	y_{t}(t) = C{\cal L}^{-1}\!\!\left\{(sE \!-\! A)^{-1}\right\}[Ex(0) \!-\! Z\xi(0)]  \\	
& & \label{decom-2-2}
	y_{s}(t) = (CX + D\Pi) \xi(t) 	
\end{eqnarray}
These expressions make it clear that the system response $y(t)$ can be explicitly divided into two constituents, the transient response $y_{t}(t)$ and the steady-state response $y_{s}(t)$.
The transient response $y_{t}(t)$ is due to the initial states of the descriptor system $ \mathbf{\Sigma}_{p} $ and those of the IGS $ \mathbf{\Sigma}_{s} $, that decays exponentially to zero with time increments, provided that the descriptor system $ \mathbf{\Sigma}_{p} $ is stable; while the steady-state response $y_{s}(t)$ has the same dynamics as the IGS $ \mathbf{\Sigma}_{s} $, noting that the matrix $CX+D\Pi$ is a constant matrix, which only depends on system matrices of the descriptor system $ \mathbf{\Sigma}_{p} $ and those of the IGS $ \mathbf{\Sigma}_{s} $, while $\xi(t)$ represent the state vector of the IGS $ \mathbf{\Sigma}_{s} $.

To simplify expressions, a generalized eigenvalue of the matrix pair $(E,A)$ and an eigenvalue of the matrix $\Xi$ are sometimes, with a little abuse of terminology, called respectively an eigenvalue of the descriptor system $ \mathbf{\Sigma}_{p} $ and the IGS $ \mathbf{\Sigma}_{s} $.

To investigate properties of the steady-state response $y_{s}(t)$, let $\lambda_{r,i}$, $i=1,2,\cdots,m_{r}$ and $\lambda_{c,i}, \lambda^{*}_{c,i}$, $i=1,2,\cdots,m_{c}$, denote respectively the real and complex eigenvalues of the IGS $ \mathbf{\Sigma}_{s}$, with their associated eigenvectors respectively being $t_{r,i}$, $t_{c,i}$ and $t^{*}_{c,i}$. Note that the matrix $\Xi$ is real valued. If it has a complex eigenvalue, then the conjugate of this complex number is also its eigenvalue, with the associated eigenvectors having the same relations \cite{hj2013}, meaning that the aforementioned expressions are always feasible. In addition, it is clear from Assumption \ref{assum:4} and the dimension of the matrix $\Xi$ that $m_{r}+2m_{c}=m_{\xi}$. Moreover, define matrix $T$ as follows,
\begin{equation}
T=\left[ t_{r,1}\;\cdots\; t_{r,m_{r}}\;t_{c,1}\; t^{*}_{c,1}\; \cdots\; t_{c,m_{c}}\; t^{*}_{c,m_{c}}\right]
\label{decom-3}
\end{equation}
Furthermore, denote the vector $\Pi t_{\star,i}$ by $\overline{\pi}_{\star,i}$, in which $i=1,2,\cdots, m_{\star}$ and $\star=r,\;c$. Recall that the matrix $\Pi$ is real valued. It is clear that $\Pi t^{*}_{c,i} = \overline{\pi}^{*}_{c,i}$ is valid for each $i=1,2,\cdots, m_{c}$. This implies that the following equality is valid,
\begin{equation}
\Pi T=\left[ \overline{\pi}_{r,1}\;\cdots\; \overline{\pi}_{r,m_{r}}\;\overline{\pi}_{c,1}\; \overline{\pi}^{*}_{c,1}\; \cdots\; \overline{\pi}_{c,m_{c}}\; \overline{\pi}^{*}_{c,m_{c}}\right]
\label{decom-4}
\end{equation}

With these matrix and vector definitions, a connection can be established between the TFM $H(s)$ of the descriptor system $\mathbf{\Sigma}_{p}$ and the matrix $CX+D\Pi$ that is associated only with its steady-state response\footnote[1]{A similar relation has been established in \cite{Astolfi2010} for a SISO LTI system with a standard state-space model, under the condition that the IGS is observable, and its minimal polynomial coincides with its characteristic polynomial. This relation plays a pivotal role in extending moment matching based model reduction to nonlinear, time-variant systems, etc.}.

\begin{Theorem}\label{theo:1}
Assume that the matrix $T$ is invertible, and all the eigenvalues of the IGS $ \mathbf{\Sigma}_{s} $ are different from those of the  descriptor system $ \mathbf{\Sigma}_{p} $. Then the matrix $CX+D\Pi$ of Equation (\ref{decom-2-2}) can be expressed as follows,
\vspace{-0.0cm}
\begin{eqnarray}  \label{eqn:theo:1}
& & \hspace*{-1.0cm} CX \!+\! D\Pi \!\!=\!\! \left[ H(\lambda_{r,1})\overline{\pi}_{r,1}\;\cdots\; H(\lambda_{r,m_{r}})\overline{\pi}_{r,m_{r}}\;H(\lambda_{c,1})\overline{\pi}_{c,1}\; \right. \nonumber \\
& & \hspace*{-0.6cm} \left. H(\lambda^{*}_{c,1})\overline{\pi}^{*}_{c,1}\; \cdots\; H(\lambda_{c,m_{c}})\overline{\pi}_{c,m_{c}}\; H(\lambda^{*}_{c,m_{c}})\overline{\pi}^{*}_{c,m_{c}}\!\right]\!\! T^{-1}
\end{eqnarray}
\end{Theorem}

A proof of this theorem is given in Appendix I.

Note that the matrix $T$, as well as the vectors $\left.\overline{\pi}_{\star,i}\right|_{i=1}^{m_{\star}}$, $\star=r,\;c$, are completely determined by the IGS $\mathbf{\Sigma}_{s}$. Theorem \ref{theo:1} makes it clear that when the vectors $\left.t_{r,i}\right|_{i=1}^{m_{r}}$, $\left.t_{c,i}\right|_{i=1}^{m_{c}}$ and $\left.t^{*}_{c,i}\right|_{i=1}^{m_{c}}$ are linearly independent, the steady-state response of the descriptor system $\mathbf{\Sigma}_{p}$ is related only to the values of its TFM $H(s)$ at the eigenvalues of the IGS $\mathbf{\Sigma}_{s}$.

When there does not exist a set of linearly independent eigenvectors $\left.t_{r,i}\right|_{i=1}^{m_{r}}$, $\left.t_{c,i}\right|_{i=1}^{m_{c}}$ and $\left.t^{*}_{c,i}\right|_{i=1}^{m_{c}}$, the matrix $\Xi$ can no longer be similar to a diagonal matrix. But it generally has a Jordan canonical form \cite{gv1989,hj2013}. In such a case, similar arguments as those in the proof of Theorem \ref{theo:1} show that, not only the value of the TFM, but also the value of its derivatives with respect to the Laplace variable $s$, are reflected in the steady-state response of the descriptor system $\mathbf{\Sigma}_{p}$.

To illustrate this observation, Appendix II gives a detailed discussion for the case when the matrix $\Xi$ has only one real eigenvalue $\lambda_{r}$, and its geometric and algebraic multiplicities being $1$ and $m_{\xi}$ respectively, as well as that this eigenvalue is distinct from each eigenvalue of the descriptor system $\mathbf{\Sigma}_{p}$. Both the conclusions and the derivations are the same when the matrix $\Xi$ has several different real/complex eigenvalues, whose geometric and algebraic multiplicities are different. The details are omitted to avoid complicated expressions.

The situations become complicated when the descriptor system $\mathbf{\Sigma}_{p}$ and the IGS $\mathbf{\Sigma}_{s}$ share some eigenvalues. In this case, Equation (\ref{decom-1}) may not have a solution. In open loop system identification, however, this case may happen very rarely, noting that the descriptor system $\mathbf{\Sigma}_{p}$ must have some stability margins to enable an identification experiment, while a probing signal $u(t)$ that decays too fast in magnitude may make the experimental data not sufficiently informative \cite{Ljung1999,ps2001}. It therefore appears safe to claim that the assumptions adopted in Theorem \ref{theo:1} are not very restrictive.

These observations also reveal that through an appropriate design of the probing signal $u(t)$, it is in principle possible to get an estimate for the value of the TFM of the descriptor system $\mathbf{\Sigma}_{p}$ at almost every interested point\footnote[2]{Some technical issues may arise when this point is too far away from the imaginary axis, as it may make a data set not sufficiently informative due to the fast magnitude decrement of the probing signal, or lead to a fast magnitude increment of the probing signal that can hardly be realized in an actual identification experiment. These need further investigations.}, as well as its derivatives with respect to the Laplace variable $s$. The associate estimation problem is discussed in detail in the next section, as the first step towards settling the aforementioned parametric estimation problem.

In addition to these, Theorem \ref{theo:1} can also be rewritten in a form with only real numbers, and applied to estimating a right tangential interpolation condition of a TFM, at almost every point of the complex plane and along an arbitrary interested direction. This condition is extensively utilized in constructing a rational matrix valued function (MVF), identifying a Port-Hamiltonian system and its data-driven model reduction, etc., but it is still not clear how to get it directly from experimental data \cite{abg2020,bgr1990,bgd2020,msa2024}.

More precisely, when the eigenvalues of the matrix $\Xi$ are different from each other, it can be assumed without any loss of generality that
\begin{equation}
\Xi = {\it diag}\!\left\{ \left.\lambda_{r,i}\right|_{i=1}^{m_{r}}, \left.\left[\begin{array}{rr} \sigma_{i} & -\omega_{i} \\ \omega_{i} & \sigma_{i} \end{array}\right]\right|_{i=1}^{m_{c}} \right\}
\label{eqn:rd-1}
\end{equation}
in which all the numbers $\lambda_{r,i}|_{i=1}^{m_{r}}$, $\sigma_{i}|_{i=1}^{m_{c}}$ and $\omega_{i}|_{i=1}^{m_{c}}$ are  real numbers, and  $\omega_{i} \neq 0$. Represent correspondingly the initial condition vector $\xi(0)$ of the IGS $\mathbf{\Sigma}_{s}$ and its output matrix $\Pi$ respectively as
\begin{eqnarray}
& &\hspace*{-1.1cm} \xi(0) = \left[\xi_{r,1}(0)\;\;\cdots\;\; \xi_{r,m_{r}}(0)\; \;\xi^{[r]}_{c,1}(0) \right. \nonumber\\
& & \hspace*{1.5cm} \left. \xi^{[i]}_{c,1}(0)\;\; \cdots \;\; \xi^{[r]}_{c,m_{c}}(0)\;\; \xi^{[i]}_{c,m_{c}}(0) \right]^{T}
\label{eqn:rd-2} \\
& &\hspace*{-1.1cm}  \Pi = \left[\pi_{r,1}\;\;\cdots\;\; \pi_{r,m_{r}}\;\; \pi^{[r]}_{c,1}\;\; \pi^{[i]}_{c,1}\;\; \cdots \;\; \pi^{[r]}_{c,m_{c}}\;\; \pi^{[i]}_{c,m_{c}} \right]
\label{eqn:rd-3}
\end{eqnarray}
with $\xi_{r,i}(0)|_{i=1}^{m_{r}}$, $\xi^{[r]}_{c,i}(0)|_{i=1}^{m_{c}}$ and $\xi^{[i]}_{c,i}(0)|_{i=1}^{m_{c}}$ being real numbers, while $\pi_{r,i}|_{i=1}^{m_{r}}$, $\pi^{[r]}_{c,i}|_{i=1}^{m_{c}}$ and $\pi^{[i]}_{c,i}|_{i=1}^{m_{c}}$ being real vectors. Here the superscripts $[r]$ and $[i]$ may cause some confusions, but they are clearly related closely to the complex eigenvalues of the matrix $\Xi$. In addition, merits of taking these superscripts can be understood without significant difficulties from the next corollary. Actually, for each $i=1,2,\cdots,m_{c}$, $\sigma_{i}$ and $\omega_{i}$ are respectively the real and imaginary parts of a complex eigenvalue of the system matrix $\Xi$, while $\pi^{[r]}_{c,i} + j \pi^{[i]}_{c,i}$ can be regarded as a right tangential interpolation direction of the TFM $H(s)$ at $s=\sigma_{i} - j\omega_{i}$. In addition, $\xi^{[r]}_{c,i}(0) \pm j \xi^{[i]}_{c,i}(0)$ becomes the initial value of a state, after a state transformation of the IGS $\mathbf{\Sigma}_{s}$ using the transformation matrix $T$ defined by Equation (\ref{app-66}).

On the basis of these representations and Theorem \ref{theo:1}, the following conclusions can be reached for the steady-state response $y_{s}(t)$ of the descriptor system $\mathbf{\Sigma}_{p}$, while their proof is deferred to Appendix I.

\newtheorem{Corollary}{Corollary}
\begin{Corollary}
Assume that the initial condition vector and the system matrices of the IGS $\mathbf{\Sigma}_{s}$ take respectively the forms of Equations (\ref{eqn:rd-1})-(\ref{eqn:rd-3}). Moreover, assume that System $\mathbf{\Sigma}_{p}$ and $\mathbf{\Sigma}_{s}$ do not share any eigenvalue. Then the steady-state response $y_{s}(t)$ of the descriptor system $\mathbf{\Sigma}_{p}$ can be expressed as
\begin{eqnarray}  \label{eqn:coro:1}
& & \hspace*{-0.9cm} y_{s}(t) \!\!=\!\! \sum_{i=1}^{m_{r}}\left\{ \left(H(\lambda_{r,i}){\pi}_{r,1}\right)\times \left( e^{\lambda_{r,i}t}\xi_{r,i}(0) \right)\right\} +  \nonumber \\
& & \hspace*{-0.0cm} \sum_{i=1}^{m_{c}}\!\left\{\!\!\left(\!\! \left[ \! H^{[r]}(\sigma_{i} \!+\! j\omega_{i})\;\; H^{[i]}(\sigma_{i} \!+\! j\omega_{i}) \!\right] \!\!\!\left[\!\!\! \begin{array}{rr} {\pi}^{[r]}_{c,i}  & {\pi}^{[i]}_{c,i} \\ {\pi}^{[i]}_{c,i} & -{\pi}^{[r]}_{c,i}\end{array} \!\!\!\right]\!\right)\!\!\times \right. \nonumber \\
& & \hspace*{0.6cm} \left.
\left(\!\! e^{\sigma_{i}t} \!\!\left[\!\!\begin{array}{rr} cos(\omega_{i}t)  & -sin(\omega_{i}t) \\ sin(\omega_{i}t) & cos(\omega_{i}t) \end{array}\!\!\right]\!\!
\left[\!\!\begin{array}{c} {\xi}^{[r]}_{c,i}(0) \\ {\xi}^{[i]}_{c,i}(0) \end{array}\!\!\right]\!\right)\!\!\right\}
\end{eqnarray}
\label{coro:1}
\end{Corollary}

Note that the TFM $H(s)$ is a real and rational MVF. It can therefore be directly declared that $H(\sigma_{i} \!-\! j\omega_{i})=H^{\star}(\sigma_{i} \!+\! j\omega_{i})$, which further leads to the following relations.
\begin{eqnarray}
& & H(\sigma_{i} \!-\! j\omega_{i})\left({\pi}^{[r]}_{c,i} + j{\pi}^{[i]}_{c,i}\right) \nonumber\\
&=& \left[H^{[r]}(\sigma_{i} \!+\! j\omega_{i}) \!-\! jH^{[i]}(\sigma_{i} \!+\! j\omega_{i})\right] \!\! \left({\pi}^{[r]}_{c,i} + j{\pi}^{[i]}_{c,i}\right) \nonumber\\
&=&
\left[H^{[r]}(\sigma_{i} \!+\! j\omega_{i}){\pi}^{[r]}_{c,i} + H^{[i]}(\sigma_{i} \!+\! j\omega_{i}){\pi}^{[i]}_{c,i}\right]+  \nonumber\\
& &
+ j \left[H^{[r]}(\sigma_{i} \!+\! j\omega_{i}){\pi}^{[i]}_{c,i} - H^{[i]}(\sigma_{i} \!+\! j\omega_{i}){\pi}^{[r]}_{c,i}\right]
\label{eqn:rd-4}
\end{eqnarray}

From Equations (\ref{eqn:coro:1}) and (\ref{eqn:rd-4}), it is clear that except the eigenvalues of the descriptor system $\mathbf{\Sigma}_{p}$, for any other $\lambda \in  \mathbb{C}$ and an arbitrary complex vector $\pi \in  \mathbb{C}^{m_{u}}$, through appropriately selecting the initial conditions of the IGS $\mathbf{\Sigma}_{s}$, as well as its system matrices $\Xi$ and $\Pi$, the value of $H(\lambda)\pi$ can in principle be estimated from its measured outputs. In systems and control theories, as well as in operator theories, etc., a value like this is usually called a right tangential interpolation condition of a MVF \cite{abg2020,bgr1990,bgd2020,msa2024}. In other words, Corollary \ref{coro:1} establishes an explicit relation between steady-state response of the descriptor system $\mathbf{\Sigma}_{p}$ and its right interpolation conditions, which enables their estimations from experimental data. More specifically, an estimation procedure can be developed which is similar to the nonparametric estimation in the following section.

\section{Parametric and Nonparametric Estimation}\label{sec:par-est}

A relation is established in the previous section between the steady-state response of the descriptor system $\mathbf{\Sigma}_{p}$ and its TFM. In addition to the value of the TFM, its derivatives and right tangential interpolation conditions, this relation makes it also feasible to estimate efficiently parameters $\theta_{i}|_{i=1}^{m_{\theta}}$ from measured system outputs. In this section, these two estimation problems are investigated, with the nonparametric estimation as the first step towards the parametric estimation.

To investigate these parameter identification problems, the following TFMs are at first defined, which are independent of the parameters $\theta_{i}|_{i=1}^{m_{\theta}}$, and can be completely determined from available knowledge of the descriptor system $\mathbf{\Sigma}_{p}$,
\begin{eqnarray}\label{plant-5}
\hspace*{-1cm}		\left[\begin{array}{cc}
			G_{yv}(s) & G_{yu}(s)\\
			G_{zv}(s) & G_{zu}(s)\end{array}\right]\!\!\!\!
&=& \!\!\!\!\left[\begin{array}{cc}
			D_{yv} & D_{yu}\\
			D_{zv} & D_{zu}\end{array}\right]+
\left[\begin{array}{c}
			C_{yx}\\
			C_{zx}\end{array}\right]\times\nonumber\\
		& & \hspace*{-1cm}\!\!\!\! (s E-A_{xx})^{-1}\!\!
\left[\begin{array}{cc}
				B_{xv} & B_{xu}\end{array}\right]
\end{eqnarray}

With these TFMs, direct algebraic manipulations show that the TFM $ H(s,\theta) $ of the descriptor system $\mathbf{\Sigma}_{p}$  can be expressed as follows  \cite{zy2022,zly2024},
\begin{eqnarray}
H(s,\theta)\!\!\!\! &=&\!\!\!\! C(\theta)\left[ sE-A(\theta) \right]^{-1}B(\theta) + D(\theta) \nonumber\\
		&=&\!\!\!\! G_{yu}(s)+G_{yv}(s)P(\theta)\times\nonumber\\
		& & \hspace*{1cm} \left[ I_{m_{z}}-G_{zv}(s)P(\theta)\right]^{-1}\!\! G_{zu}(s)
\label{decom-5}
\end{eqnarray}

It has been clarified in \cite{zy2022} and \cite{zly2024} that under Assumptions \ref{assum:1} and \ref{assum:2},  all the TFMs $H(s,\theta)$ and $G_{\star\sharp}(s)$ with $\star = y,z$, $\sharp = u,v$, are well-defined for every $ \theta \in \mathbf{\Theta}$.

On the other hand, from Lemma \ref{lemma:2} and Theorem \ref{theo:1}, it is clear that when Assumptions \ref{assum:1} - \ref{assum:5} are simultaneously satisfied, the measured outputs of the descriptor system $\mathbf{\Sigma}_{p}$  can be expressed as
\begin{eqnarray}  \label{eqn:out-meas-1}
& & \hspace*{-0.8cm} y_{m}(t_{k}) \!=\! \left[ H(\lambda_{r,1},\theta)\overline{\pi}_{r,1}\;\cdots\; H(\lambda_{r,m_{r}},\theta)\overline{\pi}_{r,m_{r}}\;H(\lambda_{c,1},\theta)\overline{\pi}_{c,1} \right. \nonumber \\
& & \hspace*{-0.2cm} \left. H(\lambda^{*}_{c,1},\theta)\overline{\pi}^{*}_{c,1}\; \cdots\; H(\lambda_{c,m_{c}},\theta)\overline{\pi}_{c,m_{c}}\; H(\lambda^{*}_{c,m_{c}},\theta)\overline{\pi}^{*}_{c,m_{c}}\right]\!\!\times  \nonumber \\
& & \hspace*{2.4cm}T^{-1}\xi(t_{k}) + y_{t}(t_{k}) + n(t_{k})
\end{eqnarray}

Recall that $\xi(t)$ stands for the state vector of the IGS $\mathbf{\Sigma}_{s}$. Its value is in general available for each sampling time instant. On the other hand, note that both the parameter vector $\theta$ and the system output measurements $y_{m}(t_{k})|_{k=1}^{N}$ are real valued, while the matrix $T$ is in general complex valued. To establish a relation that is more convenient for estimation algorithm developments, between system output measurements $y_{m}(t_{k})|_{k=1}^{N}$ and the parameter vector $\theta$, it appears necessary to distinguish the real valued elements of the vector $T^{-1}\xi(t_{k})$ from its complex valued elements. From the definition of the matrix $T$ and the fact that the vector $\xi(t_{k})$ is real valued, the following expressions can be straightforwardly established,
\begin{eqnarray}
& & \hspace*{-0.8cm} T^{-1}\xi(t_{k}) = \left[ \overline{\xi}_{r,1}(t_{k})\;\; \cdots\;\; \overline{\xi}_{r,m_{r}}(t_{k})\;\; \overline{\xi}_{c,1}(t_{k}) \right. \nonumber \\
& & \hspace*{1.8cm} \left. \overline{\xi}^{*}_{c,1}(t_{k})\;\; \cdots\;\; \overline{\xi}_{c,m_{c}}(t_{k})\;\; \overline{\xi}^{*}_{c,m_{c}}(t_{k})\right]^{T}
\label{eqn:est-4}
\end{eqnarray}
in which each of $\overline{\xi}_{r,i}(t_{k})|_{i=1}^{m_{r}}$ is a real valued scalar, while each of $\overline{\xi}_{c,i}(t_{k})|_{i=1}^{m_{c}}$ is a complex valued scalar.

Define a TFM $\overline{H}(s,\theta)$, as well as vectors $\overline{y}_{m}(t_{k})$, $\overline{u}_{\star,i}(t_{k})$ and matrices $\overline{H}_{\star,i}(\theta)$, with $i=1,2,\cdots,m_{\star}$ and $\star=r,c$, respectively as follows,
\begin{eqnarray*}
& &\hspace*{-0.8cm}
\overline{H}(s,\theta) \!=\!
G_{yv}(s)P(\theta)\!\left[ \! I_{m_{z}} \!\!\!-\! G_{zv}(s)P(\theta) \right]^{\!-\!1} \\
& &\hspace*{-0.8cm} \overline{y}_{m}(t_{k}) = y_{m}(t_{k})-\sum_{i=1}^{m_{r}}\left[\overline{\xi}_{r,i}(t_{k})G_{yu}(\lambda_{r,i})\overline{\pi}_{r,i}\right]
- \\
& &\hspace*{-0.0cm} \sum_{i=1}^{m_{c}}\left[\overline{\xi}_{c,i}(t_{k})G_{yu}(\lambda_{c,i})\overline{\pi}_{c,i} + \overline{\xi}^{*}_{c,i}(t_{k})G_{yu}(\lambda^{*}_{c,i})\overline{\pi}^{*}_{c,i}\right]  \\
& &\hspace*{-0.8cm} \overline{u}_{\star,i}(t_{k}) = \overline{\xi}_{\star,i}(t_{k})G_{zu}(\lambda_{\star,i})\overline{\pi}_{\star,i}, \hspace{0.2cm}
\overline{H}_{\star,i}(\theta) \!=\! \overline{H}(\lambda_{\star,i},\theta)
\end{eqnarray*}
Clearly, for each $i=1,2,\cdots,m_{\star}$ and $\star=r,c$, the values of the vectors $\overline{y}_{m}(t_{k})$ and $\overline{u}_{\star,i}(t_{k})$ can be straightforwardly computed from the available system output measurements and the available information about the IGS $\mathbf{\Sigma}_{s}$ and the descriptor system $\mathbf{\Sigma}_{p}$.

Note that the IGS $\mathbf{\Sigma}_{s}$ and the descriptor system $\mathbf{\Sigma}_{p}$ have a cascade connection. It is clear from Equation (\ref{decom-5}) that $H(s,\theta) \!=\! G_{yu}(s) + \overline{H}(s,\theta)G_{zu}(s)$, meaning that
the output $y(t)$ of the descriptor system $\mathbf{\Sigma}_{p}$ can be expressed as the summation of the output of the TFM $G_{yu}(s)$ under the stimulus of the output of the IGS $\mathbf{\Sigma}_{s}$, and that of the TFM $\overline{H}(s,\theta)$ under the stimulus of the output of the TFM $G_{zu}(s)$ that is again stimulated by the output of the IGS $\mathbf{\Sigma}_{s}$. And so are their steady-state responses. On the other hand, note that in the associated expression for the TFM $H(s,\theta)$, only $\overline{H}(s,\theta)$ depends on the parameter vector $\theta$. The introduction of the above $\overline{y}_{m}(t_{k})$ can be regarded as getting measurements for the steady-state response of the TFM $\overline{H}(s,\theta)$, which is more convenient in settling the parametric estimation problem.

On the basis of Equation (\ref{decom-5}), Equation (\ref{eqn:out-meas-1}) can be equivalently rewritten as follows, using the above matrix and vector definitions,
\begin{eqnarray}
& &\hspace*{-1.0cm} \overline{y}_{m}(t_{k}) = \sum_{i=1}^{m_{r}}\left[\overline{H}_{r,i}(\theta)\overline{u}_{r,i}(t_{k})\right]
+ \sum_{i=1}^{m_{c}}\left[\overline{H}_{c,i}(\theta)\overline{u}_{c,i}(t_{k}) + \right.\nonumber\\
& &\hspace*{1.6cm} \left. \overline{H}^{*}_{c,i}(\theta)\overline{u}^{*}_{c,i}(t_{k})\right] + y_{t}(t_{k}) + n(t_{k})
\label{eqn:out-meas-1-x}
\end{eqnarray}

For each $i=1,2,\cdots,m_{c}$, represent the complex vector $\overline{u}_{c,i}(t_{k})$ and the complex matrix $\overline{H}_{c,i}(\theta)$ respectively by their real and imaginary parts. That is,
\begin{displaymath}
\overline{u}_{c,i}(t_{k}) \!=\! \overline{u}^{[r]}_{c,i}(t_{k}) + j\overline{u}^{[i]}_{c,i}(t_{k}), \hspace{0.20cm}
\overline{H}_{c,i}(\theta) \!=\! \overline{H}^{[r]}_{c,i}(\theta) + j\overline{H}^{[i]}_{c,i}(\theta)
\end{displaymath}
Then Equation (\ref{eqn:out-meas-1-x}) can be further equivalently expressed as follows, which is more convenient for investigating the parameter estimation problem, noting that all the involved vectors and matrices in this equation are now real valued.
\begin{eqnarray}
& &\hspace*{-1.0cm} \overline{y}_{m}(t_{k}) = \sum_{i=1}^{m_{r}}\left[\overline{H}_{r,i}(\theta)\overline{u}_{r,i}(t_{k})\right]
+ 2\sum_{i=1}^{m_{c}}\left[\overline{H}^{[r]}_{c,i}(\theta)\overline{u}^{[r]}_{c,i}(t_{k}) - \right.\nonumber\\
& &\hspace*{1.6cm} \left. \overline{H}^{[i]}_{c,i}(\theta)\overline{u}^{[i]}_{c,i}(t_{k})\right] + y_{t}(t_{k}) + n(t_{k})
\label{eqn:out-meas-2}
\end{eqnarray}

Define matrix $\overline{H}(\theta)$ and vector $\overline{u}(t_{k})$ respectively as follows
\begin{eqnarray*}
& & \hspace*{-0.3cm}\overline{H}(\theta) \!=\! \left[{\it row}\!\!\left\{\!\!\left.\overline{H}_{r,i}(\theta)\right|_{i=1}^{m_{r}}\!\!\right\},
{\it row}\!\!\left\{\!\left[\overline{H}^{[r]}_{c,i}(\theta),\;\; \overline{H}^{[i]}_{c,i}(\theta)\right]_{i=1}^{m_{c}}\!\!\right\}\right] \nonumber\\
& & \hspace*{-0.3cm}
\overline{u}(t_{k})\!=\!
{\it col}\!\!\left\{\!\! {\it col}\!\!\left\{\!\!\left.\overline{u}_{r,i}(t_{k})\right|_{i=1}^{m_{r}}\right\},
2\,{\it col}\!\!\left\{\!\left.\!\left[\!\!\!\begin{array}{r} \overline{u}^{[r]}_{c,i}(t_{k})\\ -\overline{u}^{[i]}_{c,i}(t_{k}) \end{array}\!\!\right]\!\right|_{i=1}^{m_{c}}\!\!\right\}\!\!\right\}
\end{eqnarray*}
Moreover, define matrices $\overline{Y}_{\!\!m}(t_{1:N})$, ${Y}_{t}(t_{1:N})$, $\overline{U}(t_{1:N})$ and ${N}(t_{1:N})$ respectively as follows
\begin{eqnarray*}
& & \hspace*{-0.3cm} \overline{Y}_{\!\!m}(t_{1:N}) = \left[\!\!\begin{array}{cccc}
\overline{y}_{m}(t_{1}) \;\; \overline{y}_{m}(t_{2}) \;\;
\cdots \;\;
\overline{y}_{m}(t_{N}) \end{array}\!\!\right] \nonumber\\
& & \hspace*{-0.3cm}
{Y}_{t}(t_{1:N})= \left[\!\!\begin{array}{cccc}
y_{t}(t_{1}) \;\;
y_{t}(t_{2}) \;\;
\cdots \;\;
y_{t}(t_{N}) \end{array}\!\!\right]
\nonumber\\
& & \hspace*{-0.3cm}
\overline{U}(t_{1:N}) = \left[\!\!\begin{array}{cccc}\overline{u}(t_{1}) \;\;
\overline{u}(t_{2})
\cdots \;\;
\overline{u}(t_{N}) \end{array}\!\!\right]
\nonumber\\
& & \hspace*{-0.3cm}
{N}(t_{1:N}) = \left[\!\!\begin{array}{cccc}
n(t_{1}) \;\;
n(t_{2}) \;\;
\cdots \;\;
n(t_{N}) \end{array}\!\!\right]
\end{eqnarray*}
Then from Equation (\ref{eqn:out-meas-2}), it is immediate that the following equality is satisfied by the measured system output data
\begin{equation}
\hspace*{-0.2cm} \overline{Y}_{\!\!m}(t_{1:N}) \!=\! \overline{H}(\theta) \overline{U}(t_{1:N}) \!+\!
{Y}_{t}(t_{1:N}) \!+\! {N}(t_{1:N})
\label{eqn:est-3}
\end{equation}

When the matrix $\overline{U}(t_{1:N})$ is of full row rank (FRR), the least squares based estimate for the matrix $\overline{H}(\theta)$, denote it by $\widehat{\overline{H}}(\theta)$, is given by
\begin{equation}
\hspace*{-0.25cm} \widehat{\overline{H}}(\theta) \!=\!
\overline{Y}_{\!\!m}(t_{1:N}) \overline{U}^{T}\!(t_{1:N}) \left(\!
\overline{U}(t_{1:N})\overline{U}^{T}\!(t_{1:N})
\!\right)^{\!\!-1}
\label{eqn:est-3-a}
\end{equation}

Through vectorizing both sides of Equation (\ref{eqn:est-3}), a relation similar to the following Equation (\ref{app-39}) can be established between the measured system outputs and the matrix $\overline{H}(\theta)$. With this relation, the case can be dealt with in which the composite disturbances $n(t)$ of different sampling instants are correlated, and a priori information about their statistical distributions, if available, can also be taken into account in developing an estimation procedure. Due to space considerations, these objectives are not pursued in this paper.

Using similar procedures, an estimate can also be obtained respectively for the derivatives of the TFM of the descriptor system $\mathbf{\Sigma}_{p}$ and its right tangential interpolations. These quantities are widely used in model reduction and system identification, etc. \cite{abg2020,Astolfi2010,bgr1990,bgd2020,msa2024}, and can also be used in settling the formulated parameter estimation problem. Due to space considerations and some technical issues, detailed discussions are not included in this paper.

From the definition of the matrix $\overline{H}(\theta)$ and its above estimate, estimates can be drawn simultaneously for the matrices $\overline{H}_{r,i}(\theta)|_{i=1}^{m_{r}}$, $\overline{H}^{[r]}_{c,i}(\theta)|_{i=1}^{m_{c}}$ and $\overline{H}^{[i]}_{c,i}(\theta)|_{i=1}^{m_{c}}$. Denote their values respectively by $\widehat{\overline{H}}_{r,i}(\theta)|_{i=1}^{m_{r}}$, $\widehat{\overline{H}}^{[r]}_{c,i}(\theta)|_{i=1}^{m_{c}}$ and $\widehat{\overline{H}}^{[i]}_{c,i}(\theta)|_{i=1}^{m_{c}}$. Then on the basis of the definitions of these matrices, the following equalities can be further obtained straightforwardly
\begin{eqnarray}
& & \hspace*{-1.3cm}
\overline{H}(\lambda_{r,i},\theta) \!\!=\!\! \widehat{\overline{H}}_{r,i}(\theta) \!+\! W_{r,i}, \hspace{0.2cm} i \!=\! 1,\cdots,m_{r}
\label{eqn:est-1} \\
& & \hspace*{-1.3cm}
\overline{H}(\lambda_{c,i},\theta)
\!\!=\!\!  \widehat{\overline{H}}^{[r]}_{c,i}\!(\theta) \!+\! j\widehat{\overline{H}}^{[i]}_{c,i}\!(\theta) \!+\! W^{[r]}_{c,i} \!+\!jW^{[i]}_{c,i}, \hspace{0.1cm} i \!\!=\!\! 1,\cdots\!, m_{c}
\label{eqn:est-2}
\end{eqnarray}
Here, the matrices $W_{r,i}|_{i=1}^{m_{r}}$, $W^{[r]}_{c,i}|_{i=1}^{m_{c}}$ and $W^{[i]}_{c,i}|_{i=1}^{m_{c}}$ are all real valued, standing for estimation errors.

Recall that both the TFMs $G_{yv}(s)$ and $G_{zv}(s)$ are real MVFs. This implies that for each $i=1,\cdots,m_{r}$, both the matrices $G_{yv}(\lambda_{r,i})$ and $G_{zv}(\lambda_{r,i})$ are real valued; while for each $i=1,\cdots,m_{c}$, both the matrices $G_{yv}(\lambda_{c,i})$ and $G_{zv}(\lambda_{c,i})$ are in general with complex elements. Represent these two matrices by their real and imaginary parts respectively as
\begin{eqnarray*}
& & G_{yv}(\lambda_{c,i}) \!=\! G_{yv}^{[r]}(\lambda_{c,i}) \!+\! j G_{yv}^{[i]}(\lambda_{c,i}) \\
& & G_{zv}(\lambda_{c,i}) \!=\! G_{zv}^{[r]}(\lambda_{c,i}) \!+\! jG_{zv}^{[i]}(\lambda_{c,i})
\end{eqnarray*}
Then the following three equations can be obtained through direct algebraic manipulations from Equations (\ref{eqn:est-1}) and (\ref{eqn:est-2}), noting that the matrix $P(\theta)$ is also real valued.
\begin{eqnarray}
& & \hspace*{-1.0cm}
\left[G_{yv}(\lambda_{r,i}) + \widehat{\overline{H}}_{r,i}(\theta)G_{zv}(\lambda_{r,i})\right] P(\theta) \nonumber \\
& & \hspace*{-1.2cm} =\! \widehat{\overline{H}}_{r,i}(\theta) \!+\!  W_{r,i} \!-\! W_{r,i}G_{zv}(\lambda_{r,i})\!P(\theta)
\label{eqn:par-est-1}\\
& & \hspace*{-1.0cm}
\left[G^{[r]}_{yv}(\lambda_{c,i}) + \widehat{\overline{H}}^{[r]}_{c,i}(\theta)G^{[r]}_{zv}(\lambda_{c,i}) - \widehat{\overline{H}}^{[i]}_{c,i}(\theta)G^{[i]}_{zv}(\lambda_{c,i})\right]\!P(\theta) \nonumber \\
& & \hspace*{-1.2cm} =\! \widehat{\overline{H}}^{[r]}_{c,i}(\theta) \!+\!  W^{[r]}_{c,i} \!+\! \left[\! W^{[i]}_{c,i}G^{[i]}_{zv}(\lambda_{c,i}) \!-\!
W^{[r]}_{c,i}G^{[r]}_{zv}(\lambda_{c,i}) \!\right]\!P(\theta)
\label{eqn:par-est-2} \\
& & \hspace*{-1.0cm}
\left[G^{[i]}_{yv}(\lambda_{c,i}) + \widehat{\overline{H}}^{[r]}_{c,i}(\theta)G^{[i]}_{zv}(\lambda_{c,i}) + \widehat{\overline{H}}^{[i]}_{c,i}(\theta)G^{[r]}_{zv}(\lambda_{c,i})\right] P(\theta) \nonumber \\
& & \hspace*{-1.2cm} =\! \widehat{\overline{H}}^{[i]}_{c,i}(\theta) \!+\!  W^{[i]}_{c,i} \!-\! \left[\! W^{[r]}_{c,i}G^{[i]}_{zv}(\lambda_{c,i}) \!+\!
W^{[i]}_{c,i}G^{[r]}_{zv}(\lambda_{c,i}) \!\right]\!P(\theta)
\label{eqn:par-est-3}
\end{eqnarray}

For each $i=1,2,\cdots,m_{\xi}$ and each $k=1,2,\cdots,m_{\theta}$, define vectors $\psi_{ik}$, $\overline{h}_{i}$ and $e_{i}$ respectively as follows,
\begin{displaymath}
\hspace*{-0.15cm}
\psi_{ik}\!\!=\!\!\left\{\!\!\!\!\begin{array}{ll}
{\it vec}\left\lbrace\! \left[\!G_{yv}(\lambda_{r,i}) + \widehat{\overline{H}}_{r,i}(\theta)G_{zv}(\lambda_{r,i})\!\right]\! P_{k} \!\right\rbrace\!, &  \hspace*{-0.35cm}1 \!\leq\! i \!\leq\! m_{r}  \\
{\it vec}\left\lbrace\! \left[\!G^{[r]}_{yv}(\lambda_{c,\iota}) + \widehat{\overline{H}}^{[r]}_{c,\iota}(\theta)G^{[r]}_{zv}(\lambda_{c,\iota}) - \right.\right. &  \\
 \hspace*{0.5cm} \left.\left.\widehat{\overline{H}}^{[i]}_{c,\iota}(\theta)G^{[i]}_{zv}(\lambda_{c,\iota}) \!\right]\! P_{k} \!\right\rbrace\!,\, i\!=\!m_{r}\!+\!2\iota\!-\!1, &  \hspace*{-0.35cm}1 \!\leq\!  \iota \!\leq\! m_{c} \\
{\it vec}\left\lbrace\! \left[\!G^{[i]}_{yv}(\lambda_{c,\iota}) + \widehat{\overline{H}}^{[r]}_{c,\iota}(\theta)G^{[i]}_{zv}(\lambda_{c,\iota}) + \right.\right. & \\
 \hspace*{0.5cm} \left.\left. \widehat{\overline{H}}^{[i]}_{c,\iota}(\theta)G^{[r]}_{zv}(\lambda_{c,\iota})\!\right]\!P_{k} \!\right\rbrace\!,\;\;\; i\!=\!m_{r}\!+\!2\iota, & \hspace*{-0.35cm} 1\!\leq\! \iota \!\leq\!  m_{c}  \end{array}\right.
\end{displaymath}

\begin{displaymath}
\hspace*{-0.15cm}
\overline{h}_{i}\!=\!\left\{\!\!\!\begin{array}{ll}
{\it vec}\left\lbrace\! \widehat{\overline{H}}_{r,i}(\theta) \!\right\rbrace\!, &  \hspace*{-0.25cm}1\leq i\leq m_{r}  \\
{\it vec}\left\lbrace\! \widehat{\overline{H}}^{[r]}_{c,\iota}(\theta) \!\right\rbrace\!,\;\hspace{1cm} i\!=\!m_{r}\!+\!2\iota\!-\!1, &  \hspace*{-0.25cm}1\leq \iota\leq m_{c} \\
{\it vec}\left\lbrace\! \widehat{\overline{H}}^{[i]}_{c,\iota}(\theta) \!\right\rbrace\!,\;\hspace{1cm} i\!=\!m_{r}\!+\!2\iota & \hspace*{-0.25cm} 1\leq \iota\leq m_{c}  \end{array}\right.
\end{displaymath}

\begin{displaymath}
\hspace*{-0.15cm}
e_{i}\!\!=\!\!\left\{\!\!\!\!\begin{array}{ll}
{\it vec}\left\lbrace\! W_{r,i} \!-\! W_{r,i}G_{zv}(\lambda_{r,i})\!P(\theta)\!\right\rbrace\!, &  \hspace*{-0.35cm}1\!\leq\! i \!\leq\! m_{r}  \\
{\it vec}\left\lbrace\! W^{[r]}_{c,\iota} \!+\! \left[\! W^{[i]}_{c,\iota}G^{[i]}_{zv}(\lambda_{c,\iota}) -
\right.\right. &  \\
 \hspace*{0.5cm} \left.\left.W^{[r]}_{c,\iota}G^{[r]}_{zv}(\lambda_{c,\iota}) \!\right]\!P(\theta)\!\right\rbrace\!,\; i\!=\!m_{r}\!+\!2\iota\!-\!1, &  \hspace*{-0.35cm}1 \!\leq\! \iota \!\leq\!  m_{c} \\
{\it vec}\left\lbrace\! W^{[i]}_{c,\iota} \!-\! \left[\! W^{[r]}_{c,\iota}G^{[i]}_{zv}(\lambda_{c,\iota}) +
\right.\right. & \\
 \hspace*{0.5cm} \left.\left.W^{[i]}_{c,\iota}G^{[r]}_{zv}(\lambda_{c,\iota}) \!\right]\!P(\theta)\!\right\rbrace\!,\;\;\;\; i\!=\!m_{r}\!+\!2\iota, & \hspace*{-0.35cm} 1 \!\leq\! \iota \!\leq\! m_{c}  \end{array}\right.  \\
\end{displaymath}

Based on these definitions and the parametrization of the matrix $P(\theta)$ given by Equation (\ref{plant-4}), Equations (\ref{eqn:par-est-1})-(\ref{eqn:par-est-3}) can be equivalently expressed as follows, through vectorizing both sides of these three equations.
\begin{equation}
\Psi \theta = \overline{h} + e
\label{eqn:par-est-4}
\end{equation}
in which
\begin{displaymath}
\hspace*{-0.4cm} \Psi \!=\! \left[\psi_{ik}\right]_{i,k=1}^{i=m_{\xi},\,k=m_{\theta}},\hspace{0.15cm} \overline{h} \!=\! {\rm\bf col}\!\!\left\{\!\!\left.\overline{h}_{i}\right|_{i=1}^{i=m_{\xi}}\!\!\right\},\hspace{0.15cm}
e \!=\!
{\rm\bf col}\!\!\left\{\!\!\left.e_{i}\right|_{i=1}^{i=m_{\xi}}\!\!\right\}
\end{displaymath}

When the matrix $\Psi$ is of FCR, the following estimate $\widehat{\theta}$ is obtained for the parameter vector $\theta$, minimizing the Euclidean distance between the vectors $\Psi \theta$ and $\overline{h}$, considering elements of the vector $e$ as independent random variables.
\begin{equation}
\widehat{\theta} = (\Psi^{T}\Psi)^{-1}\Psi^{T}\overline{h}
\label{eqn:par-est-5}
\end{equation}

In the above estimate, many important factors, such as the influences of the matrix $P(\theta)$ on the vector $e$, correlations among the  elements of the vector $e$, etc., have not been taken into account. This is mainly motivated by computation simplicities, and the properties of the estimated $\overline{H}(\theta)$ that are investigated in the next section. Briefly, when the data length $N$ is large, the estimate for  $\overline{H}(\theta)$ is in general quite close to its actual value. This makes the above estimate for $\theta$ meet many essential requirements in actual applications, such as consistence, asymptotic unbiasedness, etc. However, further efforts are required to see whether a more efficient estimate can be derived for the parameter vector $\theta$, using stochastic properties of the estimate for the matrix  $\overline{H}(\theta)$.

In summary, the estimation procedure for the parameters of the descriptor system $\mathbf{\Sigma}_{p}$ consists of the following two steps.
\begin{itemize}
\item The nonparametric estimation step. Estimate the value of the TFM $\overline{H}(s,\theta)$ at each of the eigenvalues of the IGS $\mathbf{\Sigma}_{s}$, using the measured outputs of the descriptor system $\mathbf{\Sigma}_{p}$. The estimate is given by Equation (\ref{eqn:est-3-a}).
\item The parametric estimation step. Estimate the value of the parameter vector $\theta$ with the above nonparametric estimates. The estimate is given by Equation (\ref{eqn:par-est-5}).
\end{itemize}

This two step estimation procedure has some similarities with those of \cite{hfhw2022,vtc2021,ylwv2020}, in the sense that all of them have a structure of a nonparametric estimation followed by a parametric estimation. The above estimation algorithm, however, can handle a wider class of LTI systems, and there are essentially no restrictions on the IGS system in actual applications. In addition, rather than the Markov parameters of a system, it is the value of its TFM that are estimated in the nonparametric estimation step, while the former is widely believed to be not very numerically stable \cite{Ljung1999,ps2001,ylwv2020}.

It is worthwhile to mention that the estimate {\small $\widehat{\overline{H}}(\theta)$} of Equation (\ref{eqn:est-3-a}) can be computed recursively. More precisely, denote this estimate by {\small $\widehat{\overline{H}}(\theta,N)$}, expressing explicitly its dependence on the data length $N$. Define a matrix $\Phi(N)$ as $\Phi(N)= (\!\overline{U}(t_{1:N})\overline{U}^{T}\!(t_{1:N})\!)^{\!\!-1}$.
Then from the definitions of the matrices $\overline{Y}_{\!\!m}(t_{1:N})$ and $\overline{U}(t_{1:N})$, as well as the well known matrix inversion formula \cite{gv1989,hj2013,zdg1996}, straightforward algebraic manipulations show that
\begin{eqnarray}
& & \hspace*{-0.75cm} \Phi(N) \!=\! \Phi(N\!-\!1) \!-\! \frac{[\Phi(N\!-\!1)\overline{u}(t_{N})][\Phi(N\!-\!1)\overline{u}(t_{N})]^{T}}{1 \!+\! \overline{u}^{T}(t_{N})\Phi(N\!-\!1)\overline{u}(t_{N})}  \\
& & \hspace*{-0.75cm}
\widehat{\overline{H}}(\theta,N) \!=\! \widehat{\overline{H}}(\theta,N \!-\!1) \!+\! \left[ \overline{y}_{m}(t_{N}) \!-\! \widehat{\overline{H}}(\theta,N\!-\!1) \overline{u}(t_{N})  \right]\!\times \nonumber\\
 & & \hspace*{2.80cm} \frac{[\Phi(N\!-\!1)\overline{u}(t_{N})]^{T}}{1 + \overline{u}^{T}(t_{N})\Phi(N\!-\!1)\overline{u}(t_{N})}
\end{eqnarray}

By the same token, similar recursive expressions can also be obtained for the estimate $\widehat{\theta}$ of Equation (\ref{eqn:par-est-5}). The details are omitted due to their obviousness.

\section{Properties of the Estimation Algorithm}\label{sec:par-pro}

In the above section, an estimate is derived for the parameter vector $\theta$ of the descriptor system $\mathbf{\Sigma}_{p}$. However, to get this estimate, it is necessary that the matrix $\Psi$ of Equation (\ref{eqn:par-est-4}) is of FCR, while the matrix $\overline{U}(t_{1:N})$ in Equation (\ref{eqn:est-3}) is of FRR. In system identification, a condition like these is usually called a persistent excitation condition, meaning that the system to be estimated is sufficiently stimulated, such that the collected experimental data are informative enough to guarantee that, an estimate can be uniquely determined from them \cite{Ljung1999,ps2001}.

In this section, we clarify situations under which the aforementioned two conditions can be satisfied, and investigate some important asymptotic properties of the parametric estimate $\widehat{\theta}$ and the non-parametric estimate $\widehat{\overline{H}}(\theta)$, such as their asymptotic unbiasedness, consistency, etc.

For this purpose, define a vector  $\widetilde{u}_{\star,i}(t_{k})$ with $i=1,2,\cdots,m_{\star}$ and $\star=r,c$, as follows,
\begin{displaymath}
\widetilde{u}_{\star,i}(t_{k}) = \overline{\xi}_{\star,i}(t_{k})\overline{\pi}_{\star,i}
\end{displaymath}
Moreover, define a vector $\widetilde{u}(t_{k})$ and a matrix $\widetilde{U}(t_{1:N})$ respectively as
\begin{eqnarray*}
& & \hspace*{-0.8cm}
\widetilde{u}(t_{k})\!=\!
{\it col}\!\!\left\{\!\! {\it col}\!\!\left\{\!\!\left.\widetilde{u}_{r,i}(t_{k})\right|_{i=1}^{m_{r}}\right\},
{\it col}\!\!\left\{\!\left.\left[\!\!\!\begin{array}{r} \widetilde{u}^{[r]}_{c,i}(t_{k})\\ \widetilde{u}^{[i]}_{c,i}(t_{k}) \end{array}\!\!\right]\!\right|_{i=1}^{m_{c}}\!\!\right\}\!\!\right\}   \\
& & \hspace*{-0.8cm} \widetilde{U}(t_{1:N}) = \left[\!\!\begin{array}{cccc}
\widetilde{u}(t_{1}) \;\;
\widetilde{u}(t_{2}) \;\;
\cdots \;\;
\widetilde{u}(t_{N}) \end{array}\!\!\right]
\end{eqnarray*}
In addition to these, with a little abuse of terminology, define a block diagonal real matrix $G_{zu}(\mathbf{\Sigma}_{s})$ as
\begin{eqnarray*}
& & \hspace*{-1.0cm} G_{zu}(\mathbf{\Sigma}_{s}) =
{\it diag}\!\!\left\{ {\it diag}\!\!\left\{\left. G_{zu}(\lambda_{r,i})\right|_{i=1}^{m_{r}}\right\}, \right. \nonumber\\
& & \hspace*{1.4cm} \left.
{\it diag}\!\!\!\left\{\!\!\left[\!\!\begin{array}{rr}
G_{zu}^{[r]}(\lambda_{c,i}) & -G_{zu}^{[i]}(\lambda_{c,i}) \\
-G_{zu}^{[i]}(\lambda_{c,i}) & -G_{zu}^{[r]}(\lambda_{c,i})\end{array}\right]_{i=1}^{m_{c}}\!\right\}\!\!\right\}
\end{eqnarray*}
indicating that this matrix is completely determined by the values of the TFM $G_{zu}(s)$ at the eigenvalues of the IGS $\mathbf{\Sigma}_{s}$.

With these definitions, the following conclusions are obtained for the persistent excitation of the non-parametric estimate $\widehat{\overline{H}}(\theta)$. Their proof is included in Appendix I.

\begin{Lemma}
To guarantee that the matrix $\overline{U}(t_{1:N})$ is of FRR, it is necessary that the matrix $G_{zu}(\mathbf{\Sigma}_{s})$ is of FRR. When this condition is satisfied, the matrix $\overline{U}(t_{1:N})$ is of FRR, if and only if the matrix $\left[\widetilde{U}(t_{1:N})\;\; G_{zu,r}^{\perp}(\mathbf{\Sigma}_{s})\right] $ is of FRR.
\label{lemma:3}
\end{Lemma}

Note that the matrix $\widetilde{U}(t_{1:N})$ depends only on the IGS $\mathbf{\Sigma}_{s}$ and sampling instants. The above conclusions separate requirements for identifying parameters of the descriptor system $\mathbf{\Sigma}_{p}$  into those on the IGS $\mathbf{\Sigma}_{s}$, and those on the descriptor system $\mathbf{\Sigma}_{p}$ itself, which is usually appreciated in identification experiment designs \cite{Ljung1999,ps2001}. On the other hand, note that the matrix $G_{zu}(\mathbf{\Sigma}_{s})$ is of block diagonal, and so is the matrix $G_{zu,r}^{\perp}(\mathbf{\Sigma}_{s})$, whose columns form a basis for the right null space of the matrix $G_{zu}(\mathbf{\Sigma}_{s})$. These characteristics may be helpful in reducing computational complexities in verifying the conditions of the above lemma. In addition, it can be straightforwardly shown that the matrix $G_{zu}(\mathbf{\Sigma}_{s})$ is of FRR, if and only if $G_{zu}(\lambda_{\star,i})$ is of FRR at each $i=1,2,\cdots, m_{\star}$ and $\star = r, c$.

Lemma \ref{lemma:3} also makes it clear that the parameter estimation method suggested in this paper asks that the TFM $G_{zu}(s)$ is of FNRR, recalling that a MVF is of FNRR, if and only if there exists at least one value of its variable, at which the value of this MVF is of FRR \cite{zdg1996}. This condition is required also in \cite{zy2022,zly2024} respectively for identifiability verification of an NDS and sloppiness analysis of its parameters, and restricts applicabilities of the obtained results. When this condition is not satisfied, a possible modification is that, rather than $\overline{H}_{\star,i}(\theta)$, it is $H(\lambda_{\star,i},\theta)$ that is at first estimated for each $i=1,2,\cdots, m_{\star}$ and $\star = r, c$. Then the value of the parameter vector $\theta$ is reconstructed from these estimated values. This modification, however, may lead to solving a set of bilinear equations, which is not a very easy task in general, and therefore requires some more efforts. Another approach is to include some estimates for the derivatives of the TFM $\overline{H}(s,\theta)$ in the nonparametric estimation step. This is now under studies.

From Equation (\ref{eqn:coro:1}), it is clear that when the output matrix $\Pi$ is fixed for the IGS $\mathbf{\Sigma}_{s}$, the steady-state response $y_{s}(t)$ of the descriptor system $\mathbf{\Sigma}_{p}$ varies in a restricted space, which may not be informative enough for uniquely determining $\overline{H}(\theta)$. In other words, from Corollary \ref{coro:1}, it can be understood that even when the TFM $G_{zu}(s)$ is of FNRR, conditions of Lemma \ref{lemma:3} may not be easily satisfied with only one experiment, no matter how the initial conditions and the eigenvalues of the IGS $\mathbf{\Sigma}_{s}$ are adjusted, especially when the number $m_{r}+2m_{c}$ and/or $m_{z}$ is large. This phenomenon has actually been observed in numerical simulations. A possible approach to overcome these difficulties is to directly use some right tangential interpolation conditions in an estimation of the parameter vector $\theta$, noting that these interpolation conditions can in principle be estimated from system output measurements using the aforementioned Equation (\ref{eqn:coro:1}) without satisfying conditions of Lemma \ref{lemma:3}. This is also currently under investigations.

The next lemma clarifies requirements on the descriptor system $\mathbf{\Sigma}_{p}$ for uniquely determining the parametric estimate $\widehat{\theta}$, under the condition that estimations in the first step are of a sufficiently high accuracy. Its proof is given in Appendix I.

\begin{Lemma}
Assume that the matrix $G_{zu}(\mathbf{\Sigma}_{s})$ is of FRR, and the estimation errors for $\overline{H}(\theta)$ is sufficiently small. Then the matrix $\Psi$ of Equation (\ref{eqn:par-est-4}) is of FCR, if and only if the descriptor system $\mathbf{\Sigma}_{p}$ is identifiable with its TFM value at $\lambda_{\star,i}$, in which $i=1,2,\cdots, m_{\star}$, and $\star = r, c$.
\label{lemma:4}
\end{Lemma}

Recall that identifiability is a prerequisite for parameter estimation \cite{Ljung1999,ps2001,Zhou2022,zly2024}. Lemma \ref{lemma:4} makes it clear that the conditions required by the estimation procedure suggested in the previous section are not very restrictive.

To investigate important properties like unbiasedness and consistency, etc., of the suggested estimation procedure, define a vector $e_{\overline{h}}(N)$ as \vspace{-0.0cm}
\begin{equation}
e_{\overline{h}}(N) = {\it vec}\left(\widehat{\overline{H}}(\theta) - \overline{H}(\theta)\right)
\label{eqn:pro-2}
\end{equation} \vspace{-0.0cm}
denoting estimation errors about the value of the TFM $\overline{H}(s,\theta)$ at $\lambda_{\star,i}$ with $i=1,2,\cdots,m_{\star}$, and $\star=r,c$. Then the following conclusions can be established, while their proof is deferred to Appendix I.

\begin{Theorem}
Let $f_{s}(N)$ be a positive function of the data length $N$ that increases monotonically. Assume that the matrix $G_{zu}(\mathbf{\Sigma}_{s})$ is of FRR. Moreover, assume that for each $N \geq N_{\overline{h}}$ with $N_{\overline{h}}$ being a positive integer, the following inequality is satisfied by the IGS $\mathbf{\Sigma}_{s}$, \vspace{-0.05cm}
\begin{equation}
\widetilde{U}(t_{1:N}) \widetilde{U}^{T}\!(t_{1:N}) \!\geq\! f_{s}(N) I_{m_{u}m_{\xi}}
\label{eqn:pro-1}
\end{equation}\vspace{-0.05cm}\hspace*{-0.18cm}
Then there exist $N$ independent positive numbers $\kappa_{\overline{h},e}$, $\kappa_{\overline{h},c}$ and $\kappa_{\overline{h},m}$, such that the following three inequalities are simultaneously satisfied whenever $N$ is greater than $N_{\overline{h}}$,
\begin{eqnarray}
& & \hspace*{-1.5cm}\left|\left|{\it E_{x}}\!\!\left\{e_{\overline{h}}(N)\right\}\right|\right|_{2} \leq \frac{\kappa_{\overline{h},e}}{\sqrt{f_{s}(N)}}
\label{eqn:theo5-1}  \\
& & \hspace*{-1.5cm} {\it E_{x}}\!\!\left\{\!\!\left[e_{\overline{h}}(N) \!-\!\! {\it E_{x}}\!\!\left(\!e_{\overline{h}}(N)\!\right)\!\right] \!\!\left[ e_{\overline{h}}(N) \!-\!\! {\it E_{x}}\!\!\left(\!e_{\overline{h}}(N) \!\right)\! \right]^{\!T} \!\!\right\} \nonumber\\
& & \hspace*{4.5cm} \leq\! \frac{\kappa_{\overline{h},c}\overline{\sigma}(\Sigma_{n})}{f_{s}(N)}\! I
\label{eqn:theo5-2} \\
& & \hspace*{-1.5cm} {\it E_{x}}\!\!\left\{e^{T}_{\overline{h}}(N)e_{\overline{h}}(N)\right\} \leq \frac{\kappa_{\overline{h},c}\left[m_{y}m_{z}m_{\xi}\overline{\sigma}(\Sigma_{n}) + \kappa_{\overline{h},m} \right]}{f_{s}(N)}
\label{eqn:theo5-3}
\end{eqnarray}
\label{theo:3}
\end{Theorem}

\vspace{-0.43cm}
From the definition of the matrix $\widetilde{U}(t_{1:N})$, it is clear that
\begin{eqnarray}
\widetilde{U}(t_{1:N}) \widetilde{U}^{T}\!(t_{1:N}) &=& \sum_{k=1}^{N} \widetilde{u}(t_{k}) \widetilde{u}^{T}\!(t_{k}) \nonumber \\
&\geq& \sum_{k=1}^{N-1} \widetilde{u}(t_{k}) \widetilde{u}^{T}\!(t_{k}) \nonumber \\
&=& \widetilde{U}(t_{1:N-1}) \widetilde{U}^{T}\!(t_{1:N-1})
\end{eqnarray}
meaning that the minimal eigenvalue of the matrix $ \widetilde{U}(t_{1:N}) \widetilde{U}^{T}\!(t_{1:N}) $ certainly increases monotonically. That is, the existence of the function $f_{s}(N)$ is always guaranteed, provided that the matrix $\widetilde{U}(t_{1:N_{\overline{h}}})$ is of FRR.

Replace the matrix $\widetilde{U}(t_{1:N})$ with the matrix $\overline{U}(t_{1:N})$ in the above theorem. Similar conclusions can be reached for the nonparametric estimation error $e_{\overline{h}}(N)$. The details are omitted due to their obviousness.

Note that the matrix $\Sigma_{n}$ is independent of the data length $N$, and so is its maximal singular value. The above theorem makes it clear that if $\lim_{N\rightarrow\infty}f_{s}(N) = \infty$, then the estimation bias, the mean square estimation errors, as well as the covariance matrix of the estimation error of $\overline{H}(\theta)$, all asymptotically converge to zero, provided that the TFM $G_{zu}(s)$ is FNRR. Based on studies on stochastic processes and estimation theories \cite{ct1997,Ljung1999,ps2001}, this implies that the estimate $\widehat{\overline{H}}(\theta)$ converges w.p.1 asymptotically to the actual value of the TFM $\overline{H}(s,\theta)$ at each eigenvalue of the IGS $\mathbf{\Sigma}_{s}$.

In addition, properties like asymptotic normality of the estimate $\widehat{\overline{H}}(\theta)$, etc., can also be established under some weak conditions, using similar arguments for an ordinary least squares estimate as those in \cite{Ljung1999,ps2001}. The details are omitted due to space considerations.

On the basis of these conclusions about the nonparametric estimate $\widehat{\overline{H}}(\theta)$, some attractive properties can be further established for the parametric estimate $\widehat{\theta}$, characterizing its asymptotic unbiasedness, etc. To deal with these properties, the following assumptions are introduced.

\begin{Assumption}\label{assum:6}
The set $\mathbf{\Theta}$ is bounded, and the descriptor system $\mathbf{\Sigma}_{p}$ is identifiable for each $\theta \in \mathbf{\Theta}$.
\end{Assumption}

\begin{Assumption}\label{assum:7}
Each eigenvalue of the IGS $\mathbf{\Sigma}_{s}$ is not a generalized eigenvalue of the matrix pair $(E,\, A_{xx})$.
\end{Assumption}

\begin{Assumption}\label{assum:8}
There exist positive numbers $\kappa_{s}$ and $\delta_{s}$ that are independent of the data length $N$, such that for each sufficiently large $N$, the following inequality is satisfied by the function $f_{s}(N)$ of Theorem \ref{theo:3},
\begin{equation}
f_{s}(N) \geq \kappa_{s} N^{1+\delta_{s}}
\label{eqn:theo6-0}
\end{equation}
\end{Assumption}

From Lemmas \ref{lemma:3} and \ref{lemma:4}, it is clear that the satisfaction of Assumption \ref{assum:6} is a prerequisite for guaranteeing that the parametric estimate $\widehat{\theta}$ is well-defined. On the other hand, conditions like Assumption \ref{assum:8} are widely adopted in system identification for assuring some important asymptotic properties of an estimate \cite{Ljung1999,ps2001}. In addition, from the proof of the following theorem, it can be easily understood that Assumption \ref{assum:7} is actually not necessary, but its adoption can significantly simplify the proof without sacrificing basic ideas.

\begin{Theorem}
Assume that all the conditions of Theorem \ref{theo:3}, as well as Assumptions \ref{assum:6}-\ref{assum:8} are simultaneously satisfied. Denote $\widehat{\theta} - \theta$ by $e_{\theta}(N)$. Then for an arbitrary positive number $\delta_{\theta}$ that belongs to $(0,\, 1)$, there exist a positive integer $N_{\theta}$, and $N$ independent positive numbers $\kappa_{\theta,e}$, $\kappa_{\theta,c}$ and $\kappa_{\theta,m}$, such that for every $N \geq N_{\theta}$, the following three inequalities are simultaneously satisfied with a probability at least  $1-\delta_{\theta}$,
\begin{eqnarray}
& & \hspace*{-1.40cm}\left|\left|{\it E_{x}}\!\!\left\{e_{\theta}(N)\right\}\right|\right|_{2} \leq \frac{\kappa_{\theta,e}}{\sqrt{f_{s}(N)}}
\label{eqn:theo6-1}  \\
& & \hspace*{-1.40cm} {\it E_{x}}\!\!\left\{\!\!\left[e_{\theta}(N) \!-\!\! {\it E_{x}}\!\!\left(\!e_{\theta}(N)\!\right)\!\right] \!\left[ e_{\theta}(N) \!-\!\! {\it E_{x}}\!\!\left(\!e_{\theta}(N) \!\right)\! \right]^{\!T} \!\!\right\} \!\!\leq\! \frac{\kappa_{\theta,c}\overline{\sigma}(\Sigma_{n})}{f_{s}(N)}\! I
\label{eqn:theo6-2} \\
& & \hspace*{-1.40cm} {\it E_{x}}\!\!\left\{e^{T}_{\theta}(N)e_{\theta}(N)\right\} \leq \frac{\kappa_{\theta,c}\left[m_{y}m_{z}m_{\xi}\overline{\sigma}(\Sigma_{n}) + \kappa_{\theta,m} \right]}{f_{s}(N)}
\label{eqn:theo6-3}
\end{eqnarray}
\label{theo:4}
\end{Theorem}

A proof of this theorem is provided in Appendix I.

The above theorem makes it clear that in addition to simplicities in calculation, the estimate $\widehat{\theta}$ given by Equation (\ref{eqn:par-est-5}) has also many other attractive properties, such as asymptotic unbiasedness, consistence, etc. Except for a probability that can be made close to $1$ arbitrarily, Equations (\ref{eqn:theo6-1})-(\ref{eqn:theo6-3}) take completely the same form as those of Equations (\ref{eqn:theo5-1})-(\ref{eqn:theo5-3}), meaning that properties of the nonparametric estimate $\widehat{\overline{H}}(\theta)$ are almost inherited by the parametric estimate $\widehat{\theta}$.

It is worthwhile to mention that when only asymptotic properties are concerned for the estimates, Theorems \ref{theo:3} and \ref{theo:4} make it clear that there are no necessities to assume that each sampling is performed after $\overline{t}_{s}$. From the estimation algorithm, however, it is intuitively clear from both an application viewpoint and the associated proofs that, the satisfaction of this requirement is helpful in reducing estimation errors of the suggested algorithm.

From the proof of Theorem \ref{theo:4}, it is clear that a violation of Assumption \ref{assum:7} leads to an element with an infinite magnitude in the matrix  $G_{zv}(\mathbf{\Sigma}_{s})$. From Equation (\ref{app-60}), this will put some additional constraints on the vector $w(N)$, that is constituted from estimation errors of the nonparametric estimate $\widehat{\overline{H}}(\theta)$, and therefore makes the proof lengthy and complicated, noting that all the other matrices and vectors in this equation have elements with a finite magnitude. To avoid an awkward presentation, this case is not discussed in this paper.

Recall that in system identification, experimental data must be informative \cite{Ljung1999,ps2001,zly2024}. Lemma \ref{lemma:3}, as well as Theorems \ref{theo:3} and \ref{theo:4}, clarify that while sampling is permitted to be non-uniform and slow in the formulated identification problem, it does not mean that an arbitrary sampling can lead to a high quality estimate. Further efforts are still required to reveal conditions on sampling instants, such that the estimate of Equation (\ref{eqn:par-est-5}) can meet some prescribed accuracy requirements.

\section{Numerical Simulations} \label{sec:example}

Several numerical simulations have been performed with some typical NDS models and lumped systems, in order to verify the above theoretical properties of both the nonparametric and the parametric estimates. In this section, parametric estimation results with a simple mass-damper-spring system are included, in comparisons with those based on a direct least squares based data-fitting, whose associated estimate is called DLSE afterwards for brevity. These results are representative, and able to demonstrate characteristics of the suggested estimation algorithms.

The transfer function of the adopted mass-damper-spring system is described by\footnote[3]{When mass, spring and damper are replaced respectively by inductor, capacitor and resistor, this model can also be used to describe input-output relation of a circuit, for which a descriptor form model can be established \cite{abg2020,Dai1989,Duan2010}, that takes completely the same forms as those of Equations (\ref{plant-1})-(\ref{plant-3}).}
\begin{equation}
H(s) = \frac{1.0000\times 10^{2}}{ms^{2} + \mu s + k}
\label{eqn:sim-1}
\end{equation}
in which $m$, $\mu$ and $k$ represent respectively the mass, the damper constant and the spring constant. It is assumed in the numerical simulations that $m\in [5.0000 \!\times\! 10^{-1}, 1.5000]kg$, $\mu\in [3.5000, 1.0500 \!\times\! 10]N \!\cdot\! s/m$ and $k\in [1.2500 \!\times\! 10, 3.7500 \!\times\! 10]N/m$. A factor $1.0000\times 10^{2}$ is included to represent signal amplification of a position sensor. This model is widely used in describing relations between positions of the mass and external forces added on it, and can be directly obtained from some basic physical principles.

Define the parameter vector $\theta$ as
\begin{displaymath}
\theta \!=\! \left[\!\!\begin{array}{c}\theta_{1}\\ \theta_{2} \\ \theta_{3} \end{array}\!\!\right] {\rm with}\;\; \left\{\begin{array}{l} \theta_{1} \!=\! m \!-\! 1.0000\\ \theta_{2} \!=\! \mu \!-\! 7.0000 \\ \theta_{3} \!=\! k \!-\! 2.5000\!\times\! 10 \end{array}\right.
\end{displaymath}
Then from working principles of this mass-damper-spring system, a standard state-space model can be established, in which system matrices depend on this parameter vector through an LFT \cite{zdg1996}. In addition, direct algebraic manipulations show that
\begin{eqnarray*}
& & \hspace*{-0.8cm} G_{yu}(s) =  \frac{1.0000\times 10^{2}}{s^{2} +  7.0000s + 2.5000\times 10}, \hspace{0.25cm} G_{zu}(s) = 1 \\
& & \hspace*{-0.8cm} G_{zv}(s) \!\!=\!\! \frac{-[s^{2} \;\; s\;\; 1]}{s^{2} \!+\!  7.0000s \!+\! 2.5000\!\times\! 10}, \hspace{0.10cm} G_{yv}(s) \!=\! G_{yu}(s)G_{zv}(s)
\end{eqnarray*}

As a matter of fact, it can be straightforwardly checked that with these TFMs, the transfer function $H(s)$ of Equation (\ref{eqn:sim-1}) can also be expressed as follows,
\begin{displaymath}
H(s,\theta) = G_{yu}(s) + G_{yv}(s) \theta \left( 1 - G_{zv}(s) \theta \right)^{-1} G_{zu}(s)
\end{displaymath}
which has completely the same form of the second equality in Equation (\ref{decom-5}).
In addition, $G_{zu}(s)$ is obviously of FNRR, and  estimation of $m$, $\mu$ and $k$ is equivalent to that of $\theta$.

When $m=1.0000kg$, $\mu=7.0000N \!\cdot\! s/m$ and $k=2.5000 \!\times\! 10 N/m$, which is equivalent to $\theta=0$, the bandwidth of the associated mass-damper-spring system is approximately equal to $5.8469rad/s$, meaning that the Nyquist frequency is approximately $1.8621Hz$. Taking this into account, the distance between two succeeding sampling instants is set to be a continuous random variable uniformly distributed over the interval $[0.2000,\;1.0000]s$, reflecting the nonuniform and slow sampling characteristics of the numerical simulations. In addition, an upper bound $\overline{t}_{s}$ for the system settling time is estimated as $\overline{t}_{s}=2.3258s$, with the step response of the system under these physical parameter values.

In the simulations, the composite disturbance $n(t_{k})$ is assumed to be normally distributed, with its mean and standard deviation respectively being $0$ and $\sigma$, while the system matrices and initial condition vector of the IGS $\mathbf{\Sigma}_{s}$ are set to be
\begin{eqnarray*}
& &\hspace*{-1.1cm} \Xi = {\it diag}\!\left\{\! \left[\!\!\begin{array}{rr} \sigma_{1} & \omega_{1} \\ -\omega_{1} & \sigma_{1} \end{array}\!\!\right]\!,\;\; \left[\!\!\begin{array}{rr} \sigma_{2} & \omega_{2} \\ -\omega_{2} & \sigma_{2} \end{array}\!\!\right]\! \right\}  \\
& &\hspace*{-1.1cm}  \xi(0) = \left[1.0000\;\;1.0000\;\; 1.0000\;\; 1.0000 \right]^{T}
\\
& &\hspace*{-1.1cm}  \Pi = \left[2.5000\;\; 2.5000\;\; 5.0000\;\; 5.0000\right]\times 10^{-1}
\end{eqnarray*}

\renewcommand{\thefigure}{\arabic{figure}}
\setcounter{figure}{0}
\vspace{-0.1cm}
\begin{figure}[!ht]
\vspace{-0.4cm}
\begin{center}
\includegraphics[width=3.0in]{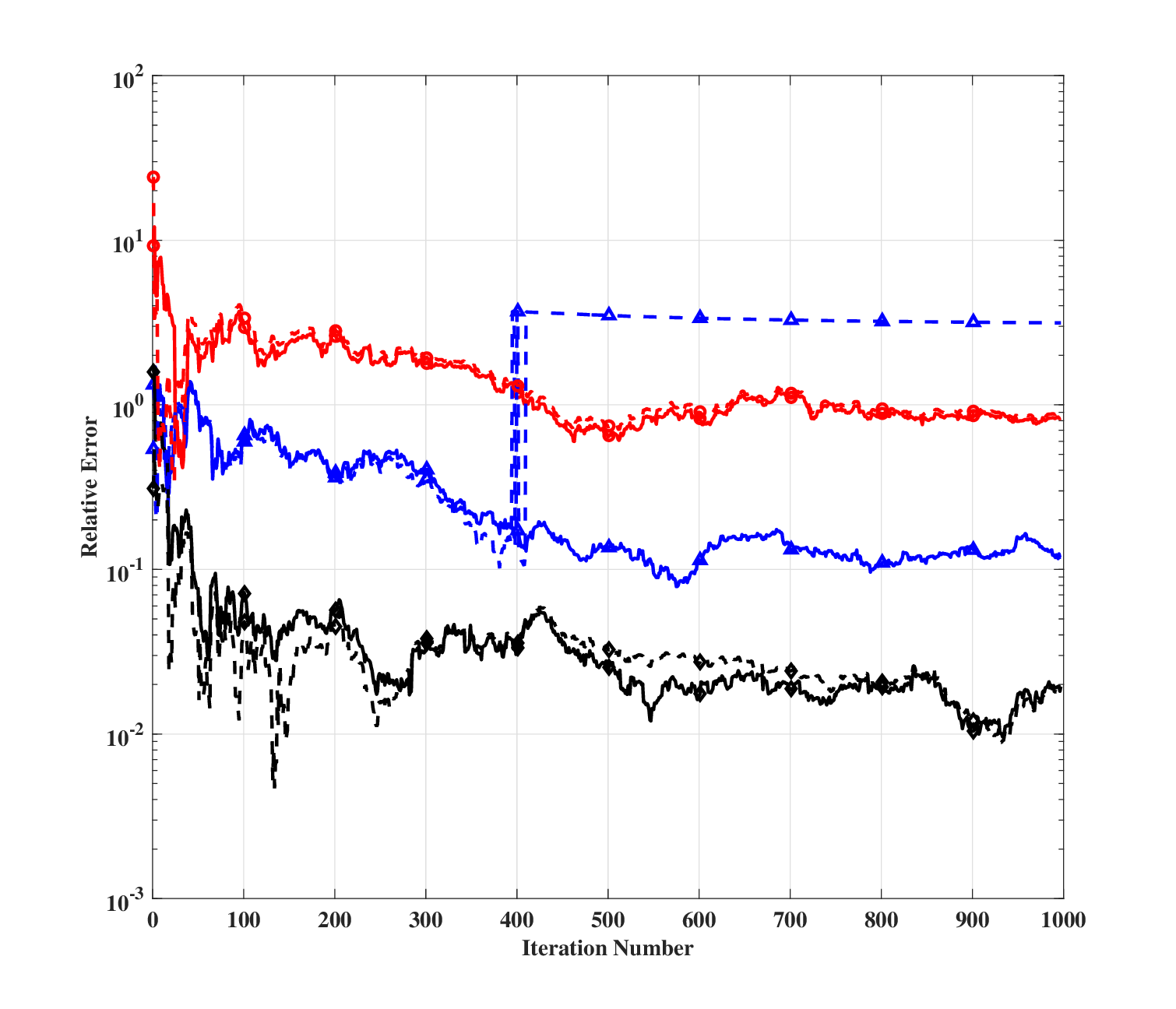}
\vspace{-0.8cm}\hspace*{5cm} \caption{Relative Error With Different Disturbance Strength.
$-\:-$: DLSE;  $-\!-$: proposed method.
$\Diamond$: $\sigma=5.0000\!\times\! 10^{-\!2}$; $\triangle$: $\sigma=2.5000\!\times\! 10^{-\!1}$; $\circ$: $\sigma=7.5000\!\times\! 10^{-\!1}$.}
\label{fig:1}
\vspace{-0.4cm}
\end{center}
\end{figure}

When the output of the above IGS $\mathbf{\Sigma}_{s}$ is added to the mass-damper-spring system, under a zero initial condition, its output $y(t,\theta)$ can be straightforwardly obtained through the inverse Laplace transformation. Specifically, the system output $y(t,\theta)$ have the following expression
\begin{eqnarray*}
y(t,\!\theta) \!\!\!\!\!\!&=&\!\!\!\!\!\! {\cal L}^{\!-\!1}\!\!\left\{\!\frac{1.0000\times 10^{2}}{(\!\theta_{1} \!\!+\!\!1.0000\!)s^{2} \!\!+\!\! (\!\theta_{2} \!\!+\!\! 7.0000\!) s \!\!+\!\! (\!\theta_{3} \!\!+\!\! 2.5000\!\!\times\!\! 10\!)} \!\times \right. \\
& &\hspace*{0.75cm} \left. \left(\!\frac{5.0000\!\!\times\!\! 10^{-\!1}(s-\sigma_{1})}{s^{2}-2\sigma_{1}s+ \omega_{1}^{2}}+ \frac{s-\sigma_{2}}{s^{2}-2\sigma_{2}s+ \omega_{2}^{2}} \!\right)     \!\!\right\}
\end{eqnarray*}

With these expressions, the DLSE is obtained through minimizing the following cost function
\begin{displaymath}
J(\theta) = \frac{1}{N+m+1}\sum_{k=-m}^{N}\left[y_{m}(t_{k}) - y(t_{k},\theta) \right]^{2}
\end{displaymath}
in which $t_{k}$ with $k \leq 0$ stands for a sampling instant before $\overline{t}_{s}$, the upper bound  of the system settling time.  The matlab nonlinear optimization M-file {\it lsqcurvefit.mat} is adopted with the default settings, and the initial estimates are randomly generated according to continuous uniform distributions with $\widehat{\theta}_{1}(0) \!\in\! [-5.0000\!\times\! 10^{-\!1},\; 5.0000\!\times\! 10^{-\!1}]$, $\widehat{\theta}_{2}(0) \!\in\! [-3.5000,\; 3.5000]$ and $\widehat{\theta}_{3}(0) \!\in\! [-1.2500\!\times\! 10,\; 1.2500\!\times\! 10]$.

In  evaluating performances of the associated parameter estimates, the Euclidean norm of a vector constituted from the relative estimation error of every individual parameter is adopted, which is defined as follows
\begin{displaymath}
E_{re}=\sqrt{\sum_{i=1}^{3}\frac{(\theta_{i}-\widehat{\theta}_{i})^{2}}{\theta_{i}^{2}}}
\end{displaymath}

More than 1,000 simulations have been performed, in which $m$, $\mu$ and $k$ are independently and randomly generated according to a continuous uniform distribution. It is observed that even for this simple system, the DLSE sometimes meets numerical stability problems, that make the optimization with {\it lsqcurvefit.mat} for the cost function $J(\theta)$ end without a result, and there are also many cases in which this  M-file ends at a local minimum.  All these problems have been successfully avoided by the suggested estimation algorithm.

\begin{figure}[!ht]
\vspace{-0.4cm}
\begin{center}
\includegraphics[width=3.0in]{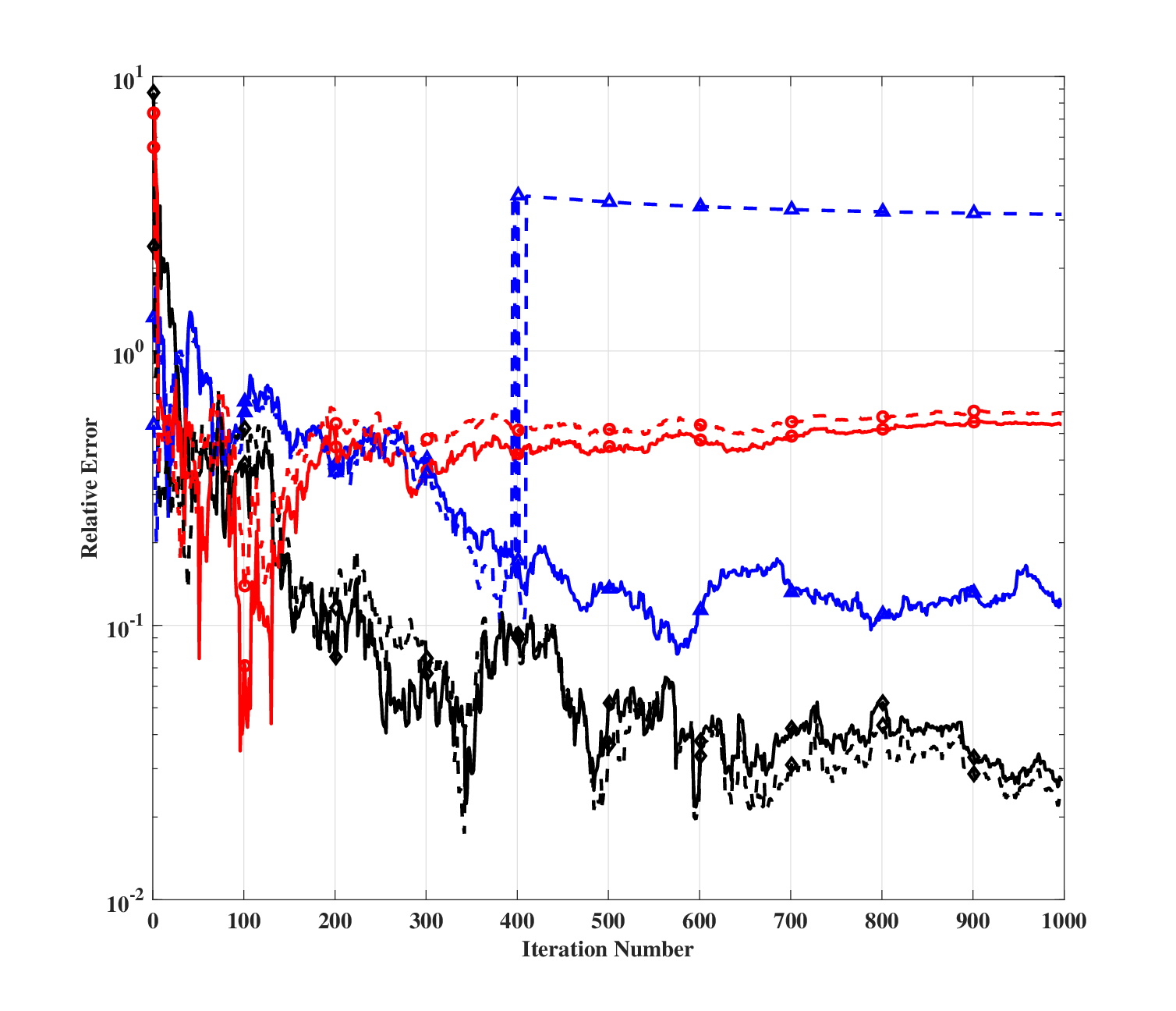}
\vspace{-0.8cm}\hspace*{5cm} \caption{Relative Error With Different Input Magnitude.
$-\:-$: DLSE;  $-\!-$: proposed method.
$\Diamond$: $\sigma_{1}=5.0000\times 10^{-3},\;\sigma_{2}=4.0000\times 10^{-3}$; $\triangle$: $\sigma_{1}=0.0000,\;\sigma_{2}=0.0000$;
$\circ$: $\sigma_{1}=-5.0000\times 10^{-3},\;\sigma_{2}=-4.0000\times 10^{-3}$.}
\label{fig:2}
\vspace{-0.4cm}
\end{center}
\end{figure}

Figures \ref{fig:1} and \ref{fig:2} give some typical results respectively with variations of the strength of the composite disturbances $n(t_{k})$, that is $\sigma$, and those of the real parts of the eigenvalues of the IGS $\mathbf{\Sigma}_{s}$, that is $\sigma_{i}$, $i=1,2$. The latter leads to magnitude variations of the input signals to the mass-damper-spring system. In the associated  simulations, the randomly generated parameter values for $m$, $\mu$ and $k$ correspond to $\theta_{1}=0.1852kg$, $\theta_{2}=0.5126N \!\cdot\! s/m $ and $\theta_{3}=6.2582N/m$, while the parameters $\omega_{i}|_{i=1}^{2}$ are selected as $\omega_{1}=3.0000rad/s$ and $\omega_{2}=4.5000rad/s$. In Figure \ref{fig:1} associated simulations, the parameters $\sigma_{i}|_{i=1}^{2}$ are fixed to be $\sigma_{1}=\sigma_{2}=0.0000$, while in Figure  \ref{fig:2} associated simulations, the parameter $\sigma$ is fixed to be $\sigma=2.5000\!\times\! 10^{-\!1}$. With these parameter values, the bandwidth and the Nyquist frequency of the associated mass-damper-spring system are approximately $6.0054rad/s$ and $1.9126Hz$ respectively, meaning that in the randomly generated sampling instants, most of the sampling rates are smaller than the Nyquist frequency.

Obviously, when the nonlinear optimization with the M-file {\it lsqcurvefit.mat} successfully reaches a global minimum, estimation accuracies  of the proposed method and the DLSE are almost the same. In addition, assigning an eigenvalue of the IGS $\mathbf{\Sigma}_{s}$ in the closed left complex plane does not appear to be a good option in estimation accuracy improvement.

\section{Conclusions}\label{sec:conclusions}

An estimation procedure is derived for parameters of a descriptor system, in which system matrices depend on parameters through an LFT. The sampling instants can be arbitrary, provided that some associated rank conditions are satisfied by the exciting signals. Several explicit relations have been established between system steady-state response and its TFM. These relations make it clear that through an appropriate input design, the values of the TFM of a descriptor system, as well as its derivatives and right tangential interpolations, can be estimated at almost every interested point. Properties like asymptotic unbiasedness, consistence, etc., have been derived for the estimate.

Further efforts include how to efficiently incorporate transient responses of the system into its parameter estimations, as well as how to more efficiently take the statistical properties in the nonparametric estimation step into account. Optimal selection of sampling instants and IGS system matrices, as well as statistical properties of the suggested estimates with finitely many data, are also of great importance. It is also interesting to see whether the FNRR condition on the TFM $G_{zu}(s)$ can be efficiently removed, through introducing some estimates of the derivatives and/or the right tangential interpolations of the TFM of the descriptor system into the nonparametric estimation step, as well as to study how the obtained results can be applied to the case in which several sets of identification experimental data are available, which are performed under distinct settings for initial states, probing signals, and sampling instants, etc.

\small
\vspace{0.25cm}
\hspace*{-0.45cm}{\rm\bf Acknowledgements.}
The author would like to thank Mr. Y. X. Ma for various discussions and help in performing numerical simulations.

\renewcommand{\labelenumi}{\rm\bf A\arabic{enumi})}

\renewcommand{\theequation}{A\arabic{equation}}
\setcounter{equation}{0}

\small
\appendices
\section*{Appendix I. Proof of Some Technical Results}

\hspace*{-0.40cm}{\rm\bf Proof of Lemma \ref{lemma:2}.} Let $\star(s)$ denote the Laplace transform of a vector valued function $\star(t)$. Take Laplace transformation of both sides of Equations (\ref{plant-1}) and (\ref{signal-1}), and substitute the 2nd equation into the first equation after the transformations, we have that
\begin{equation}
\left[\begin{array}{c}
E[sx(s)-x(0)] \\  s\xi(s)-\xi(0) \end{array}\right]
=
\left[\begin{array}{cc} A & B\Pi \\ 0 & \Xi \end{array}\right]
\left[\begin{array}{c}
x(s) \\  \xi(s) \end{array}\right]
\end{equation}
Hence
\begin{eqnarray}\label{app-1}
\hspace*{-0.5cm} \left[\!\!\!\begin{array}{c}
x(s) \\ \xi(s) \end{array}\!\!\!\right]
\!\!\!\!\!\!&=&\!\!\!\!\!\!\left[\begin{array}{cc}
sE-A & -B\Pi \\ 0 & sI-\Xi \end{array}\right]^{-1}
\left[\begin{array}{c}
Ex(0) \\  \xi(0) \end{array}\right]  \nonumber\\
&=&\!\!\!\!\!\!\left[\!\!\!\!\begin{array}{cc}
(sE\!-\!A)^{-1} & (sE\!-\!A)^{-1}B\Pi(sI\!-\!\Xi)^{-1} \\ 0 & (sI\!-\!\Xi)^{-1} \end{array}\!\!\!\!\right]
\!\!\left[\!\!\begin{array}{c}
Ex(0) \\  \xi(0) \end{array}\!\!\!\!\right]
\end{eqnarray}

When the conditions of Equation (\ref{decom-1}) are satisfied, we have that
\begin{equation}\label{app-2}
B\Pi = Z\Xi - AX = (sE-A)X - Z(sI-\Xi) 	
\end{equation}
Substitute this relation into Equation (\ref{app-1}). Direct algebraic manipulations show that
\begin{eqnarray}
\hspace*{-0.5cm}
& & \hspace*{-0.5cm}  \xi(s) = (sI\!-\!\Xi)^{-1}\xi(0) \\
& &\hspace*{-0.5cm}  x(s) = (sE\!-\!A)^{-1}Ex(0) + (sE\!-\!A)^{-1}\left[(sE-A)X - \right. \nonumber\\
& & \hspace*{2.5cm} \left. Z(sI-\Xi)\right](sI\!-\!\Xi)^{-1}\xi(0) \nonumber \\
& & \hspace*{0.2cm}       = (sE\!-\!A)^{-1}[Ex(0)-Z\xi(0)] + X\xi(s)
\end{eqnarray}

Recall that $X$ is a constant matrix. On the basis of the linearity property of Laplace transformations, we further have that
\begin{equation}\label{app-3}
x(t) = {\cal L}^{-1}\!\!\left\{ (sE\!-\!A)^{-1}\right\}[Ex(0)-Z\xi(0)] + X\xi(t)
\end{equation}
Note that both the matrices $E$ and $Z$ are time independent. The equality of Equation (\ref{decom-1-x}) is satisfied with $\overline{x}(0) = Ex(0)-Z\xi(0)$.

On the contrary, assume that there is a time independent vector $\overline{x}(0)$, together with a time independent matrix $X$, such that Equation (\ref{decom-1-x}) is satisfied at every time constant $t \geq 0$. Taking Laplace transformation of both sides of this equation, we have that
\begin{equation}\label{app-11}
	x(s) = (sE \!-\! A)^{-1}\overline{x}(0) \!+\! X \xi(s) 	
\end{equation}

Substitute Equation (\ref{app-1}) into this equation. It can be straightforwardly shown that
\begin{eqnarray}
\hspace*{-0.2cm} Ex(0)-\overline{x}(0) \!\!\!\!&=&\!\!\!\! (sE \!-\! A)X(sI \!-\! \Xi)^{-1}\xi(0) \!-\! B\Pi (sI \!-\! \Xi)^{-1}\xi(0) \nonumber \\
\!\!\!\!&=&\!\!\!\! \left[(sE \!-\! A)X \!-\! B\Pi \right](sI \!-\! \Xi)^{-1}\xi(0)
\label{app-12}
\end{eqnarray}
Note that both the vectors $\xi(0)$ and $Ex(0)-\overline{x}(0)$ do not depend on the Laplace variable $s$. It is necessary that there exists a constant matrix $Z$, such that
\begin{equation}\label{app-13}
	(sE \!-\! A)X \!-\! B\Pi \equiv Z (sI \!-\! \Xi) 	
\end{equation}
That is
\begin{equation}
	(EX \!-\! Z)s \!+\! (Z\Xi\!-\!AX \!-\! B\Pi) \equiv 0 	
\end{equation}
which is equivalent to Equation (\ref{decom-1}), and therefore completes the proof.  \hspace{\fill}$\Diamond$

\hspace*{-0.40cm}{\rm\bf Proof of Theorem \ref{theo:1}.} Define a matrix $\Lambda$ as follows,
\begin{displaymath}
\Lambda={\it diag}\!\left\{\left.\lambda_{r,i}\right|_{i=1}^{m_{r}},
\left.\left[\begin{array}{cc} \lambda_{c,i} & 0 \\ 0 & \lambda^{*}_{c,i} \end{array}\right]\right|_{i=1}^{m_{c}} \right\}
\end{displaymath}
Then under the assumption that the matrix $T$ is invertible, standard results in matrix analysis \cite{hj2013} lead to the following equality
\begin{equation}\label{app-4}
	\Xi = T \Lambda T^{-1}	
\end{equation}

Substitute this relation into Equation (\ref{decom-1}). The next equality is further obtained
\begin{equation}\label{app-5}
	 EXT \Lambda = AXT + B\Pi T	
\end{equation}

Denote the vector $X t_{\star,i}$ by $\overline{x}_{\star,i}$, in which $i=1,2,\cdots, m_{\star}$ and $\star=r,\;c$.
Note that the matrix $X$ is real valued. It is clear that for each $i=1,2,\cdots, m_{c}$, $X t^{*}_{c,i} = \overline{x}^{*}_{c,i}$, which implies that
\begin{equation}\label{app-14}
	 XT = \left[ \overline{x}_{r,1}\;\cdots\; \overline{x}_{r,m_{r}}\;\overline{x}_{c,1}\; \overline{x}^{*}_{c,1}\; \cdots\; \overline{x}_{c,m_{c}}\; \overline{x}^{*}_{c,m_{c}}\right]	
\end{equation}

On the basis of this relation, as well as the diagonal structure of the matrix $\Lambda$, and the fact that the matrices $E$, $A$ and $B$ are also real valued, Equation (\ref{app-5}) can be equivalently rewritten as follows,
\begin{eqnarray}
& & \label{app-6}
\lambda_{r,i}E\overline{x}_{r,i} = A\overline{x}_{r,i} + B \overline{\pi}_{r,i}, \hspace{0.25cm} i=1,2,\cdots,m_{r}	\\
& & \label{app-7}
\lambda_{c,i}E\overline{x}_{c,i} = A\overline{x}_{c,i} + B \overline{\pi}_{c,i}, \hspace{0.25cm} i=1,2,\cdots,m_{c}
\end{eqnarray}
which can be further expressed as
\begin{eqnarray}
& & \label{app-8}
\hspace*{-1cm} \overline{x}_{r,i} = (\lambda_{r,i}E - A)^{-1}B \overline{\pi}_{r,i}, \hspace{0.25cm} i=1,2,\cdots,m_{r}	\\
& & \label{app-9}
\hspace*{-1cm} \overline{x}_{c,i} = (\lambda_{c,i}E - A)^{-1} B \overline{\pi}_{c,i}, \hspace{0.25cm} i=1,2,\cdots,m_{c}
\end{eqnarray}
noting that the involved matrix inverses always exist, as it is assumed that for each $i=1,2,\cdots,m_{\star}$ and each $\star = r, c$, $\lambda_{\star,i}$ is not a generalized eigenvalue of the matrix pair $(E,\,A)$.

Substitute the equalities of Equations (\ref{app-8}) and (\ref{app-9}) into Equation (\ref{app-14}), and recall that all the matrices $A$, $B$, $C$ and $D$ are real valued. We have that
\begin{eqnarray} \label{app-10}
 \!\!\!\!& &\!\!\!\! CX + D\Pi  \nonumber \\
\!\!\!\!&=&\!\!\!\! \left[ CXT + D\Pi T\right] T^{-1} \nonumber \\
\!\!\!\!&=&\!\!\!\! \left[ C\left[ \overline{x}_{r,1}\;\cdots\; \overline{x}_{r,m_{r}}\;\overline{x}_{c,1}\; \overline{x}^{*}_{c,1}\; \cdots\; \overline{x}_{c,m_{c}}\; \overline{x}^{*}_{c,m_{c}}\right] + \right. \nonumber \\
& & \left. D\left[ \overline{\pi}_{r,1}\;\cdots\; \overline{\pi}_{r,m_{r}}\;\overline{\pi}_{c,1}\; \overline{\pi}^{*}_{c,1}\; \cdots\; \overline{\pi}_{c,m_{c}}
\; \overline{\pi}^{*}_{c,m_{c}}\right] \right]T^{-1} \nonumber \\
\!\!\!\!&=&\!\!\!\!
\left[{\it row}\!\left\{ \left.\left[C(\lambda_{r,i}E - A)^{-1}B+D\right]\overline{\pi}_{r,i}\right|_{i=1}^{m_{r}}\right\}\right. \nonumber \\
& &
\left.{\it row}\!\left\{ \left.\left[C(\lambda_{c,i}E - A)^{-1}B+D\right]\overline{\pi}_{c,i},\right.\right. \right.
\nonumber \\
& & \left.\left.\left.
\left[C(\lambda^{*}_{c,i}E - A)^{-1}B+D\right]\overline{\pi}^{*}_{c,i}
\right|_{i=1}^{m_{c}}\right\}
\right]T^{-1}
\end{eqnarray}

The proof can now be completed using Equation (\ref{decom-5}).
\hspace{\fill}$\Diamond$

\hspace*{-0.40cm}{\rm\bf Proof of Corollary \ref{coro:1}.} Define matrices $T$ and $\Lambda$ respectively as
\begin{eqnarray}
& &\hspace*{-1.5cm}  T = {\it diag}\!\left\{ I_{m_{r}}, \left.\left[\begin{array}{rr} 1 & 1 \\ -j & j \end{array}\right]\right|_{m_{c}\;\;{\rm blocks}} \right\}
\label{app-66} \\
& & \hspace*{-1.5cm} \Lambda = {\it diag}\!\left\{ \left.\lambda_{r,i}\right|_{i=1}^{m_{r}}, \left.\left[\begin{array}{cc} \sigma_{i}+j\omega_{i} & 0 \\ 0 & \sigma_{i}-j\omega_{i} \end{array}\right]\right|_{i=1}^{m_{c}} \right\}
\end{eqnarray}
Then it can be directly verified that the matrix $T$ is invertible. Specifically, we have that
\begin{equation}
T^{-1} = {\it diag}\!\left\{ I_{m_{r}}, \left.\frac{1}{2}\!\!\left[\!\! \begin{array}{rr} 1 & j \\ 1 & -j \end{array}\!\! \right]\right|_{m_{c}\;\;{\rm blocks}} \right\}
\end{equation}
In addition, when the system matrix $\Xi$ takes the form of Equation (\ref{eqn:rd-1}), straightforward matrix manipulations show that
\begin{equation}
\Xi T = T \Lambda
\end{equation}
which is equivalent to $\Xi  = T \Lambda T^{-1}$, meaning that $\left.\lambda_{r,i}\right|_{i=1}^{m_{r}}$ and $\left.\sigma_{i}\pm j\omega_{i}\right|_{i=1}^{m_{c}}$ exhaust all the eigenvalues of the matrix $\Xi$, with each column vector of the matrix $T$ being an associated eigenvector. Hence, the matrix $T$ defined above are consistent with that of Equation (\ref{decom-3}).

On the basis of these relations and Equation (\ref{signal-1}), as well as properties of a matrix exponential function, we have that the state vector $\xi(t)$ of the IGS $\mathbf{\Sigma}_{s}$ can be expressed as
\begin{eqnarray}
\xi(t)\!\!\!\! &=& \!\!\!\! T e^{\Lambda t} T^{-1}\xi(0) \nonumber \\
&=&\!\!\!\! T {\it diag}\!\left\{\!\! \left.e^{\lambda_{r,i}t}\right|_{i=1}^{m_{r}}, \left[ \!\!\begin{array}{c} e^{\sigma_{i}t}\!\left[cos(\omega_{i}t)+j sin(\omega_{i}t)\right]  \\ 0  \end{array}\right.\right. \nonumber \\
& & \hspace*{0.15cm}
\left.\left. \left.\begin{array}{c}  0 \\  e^{\sigma_{i}t}\!\left[cos(\omega_{i}t)-j sin(\omega_{i}t)\right] \end{array}\!\!\right]\right|_{i=1}^{m_{c}} \!\right\}T^{-1} \! \xi(0)
\end{eqnarray}

From Theorem \ref{theo:1}, when the IGS $\mathbf{\Sigma}_{s}$ and the descriptor system $\mathbf{\Sigma}_{p}$ do not share an eigenvalue, the matrix $CX+D\Pi$  can be expressed by Equation (\ref{eqn:theo:1}).
Substitute the above equation and Equation (\ref{eqn:theo:1}) into Equation (\ref{decom-2-2}). Straightforward but tedious algebraic operations lead to Equation (\ref{eqn:coro:1}), using the expressions of Equations (\ref{eqn:rd-2}) and (\ref{eqn:rd-3}) respectively for the initial condition vector $\xi(0)$ of the IGS $\mathbf{\Sigma}_{s}$ and its output matrix $\Pi$. This completes the  proof. \hspace{\fill}$\Diamond$

\hspace*{-0.40cm}{\rm\bf Proof of Lemma \ref{lemma:3}.}
From the definition of $\overline{\xi}_{\star,i}(t_{k})$, in which $i=1,2,\cdots,m_{\star}$ and $\star =r, c$, that is given by Equation (\ref{eqn:est-4}), it is clearly a scalar. Then on the basis of the definition of $\overline{u}_{\star,i}(t_{k})$ given immediately after that equation, it is straightforward that
\begin{equation}
\overline{u}_{\star,i}(t_{k}) = G_{zu}(\lambda_{\star,i}) \widetilde{u}_{\star,i}(t_{k})
\label{app-21}
\end{equation}

Therefore, the following equality is valid for every $i=1,2,\cdots,m_{c}$,
\begin{equation}
\left[\begin{array}{r}
\overline{u}^{[r]}_{c,i}(t_{k}) \\
-\overline{u}^{[i]}_{c,i}(t_{k}) \end{array}\right]
 =
\left[\begin{array}{rr}
G_{zu}^{[r]}(\lambda_{c,i}) & -G_{zu}^{[i]}(\lambda_{c,i}) \\
-G_{zu}^{[i]}(\lambda_{c,i}) & -G_{zu}^{[r]}(\lambda_{c,i}) \end{array}\right]
\left[\begin{array}{r}
\widetilde{u}^{[r]}_{c,i}(t_{k}) \\
\widetilde{u}^{[i]}_{c,i}(t_{k}) \end{array}\right]
\label{app-22}
\end{equation}
Then Equations (\ref{app-21}) and (\ref{app-22}), as well as the definitions of the matrices $G_{zu}(\mathbf{\Sigma}_{s})$, $\overline{U}(t_{1:N})$ and $\widetilde{U}(t_{1:N})$, immediately lead to
\begin{equation}
\overline{U}(t_{1:N}) = G_{zu}(\mathbf{\Sigma}_{s}) \widetilde{U}(t_{1:N})
\label{app-23}
\end{equation}

Assume that the matrix $\overline{U}(t_{1:N}) $ is of FRR. Then for an arbitrary non-zero vector $\alpha$ with an appropriate dimension, it is necessary that
\begin{equation}
\alpha^{T} \overline{U}(t_{1:N}) = \left(\alpha^{T} G_{zu}(\mathbf{\Sigma}_{s})\right) \widetilde{U}(t_{1:N}) \neq 0
\label{app-24}
\end{equation}
meaning that $\alpha^{T} G_{zu}(\mathbf{\Sigma}_{s}) \neq 0$, which is equivalent to that the matrix $G_{zu}(\mathbf{\Sigma}_{s})$ is of FRR.

On the other hand, from Equation (\ref{app-23}), we have that the matrix $\overline{U}(t_{1:N}) $ is of FRR is equivalent to that
the matrix
\begin{displaymath}
\left(G_{zu}(\mathbf{\Sigma}_{s}) \widetilde{U}(t_{1:N})\right)^{T} =  \widetilde{U}^{T}(t_{1:N}) G^{T}_{zu}(\mathbf{\Sigma}_{s})
\end{displaymath}
is of FCR. Then it can be declared from Lemma \ref{lemma:1} that the matrix $\overline{U}(t_{1:N}) $ is of FRR, if and only if the following matrix is of FCR
\begin{equation}
\left[ \begin{array}{c}
G_{zu,l}^{T,\perp}(\mathbf{\Sigma}_{s}) \\ \widetilde{U}^{T}(t_{1:N})
\end{array}\right]
\label{app-25}
\end{equation}

Note that
\begin{equation}
\left[ G_{zu,l}^{T,\perp}(\mathbf{\Sigma}_{s})G^{T}_{zu}(\mathbf{\Sigma}_{s})\right]^{T}
=G_{zu}(\mathbf{\Sigma}_{s})\left(G_{zu,l}^{T,\perp}(\mathbf{\Sigma}_{s})\right)^{T}
\label{app-26}
\end{equation}
We therefore have that
\begin{equation}
\left(G_{zu,l}^{T,\perp}(\mathbf{\Sigma}_{s})\right)^{T}
= G_{zu,r}^{\perp}(\mathbf{\Sigma}_{s})
\label{app-27}
\end{equation}
The proof can now be completed by substituting Equation (\ref{app-27}) into the transpose of the matrix of Equation (\ref{app-25}). \hspace{\fill}$\Diamond$

{\rm\bf Proof of Lemma \ref{lemma:4}.} From Equations (\ref{eqn:est-1}) and (\ref{eqn:est-2}), direct matrix manipulations show that for each $i=1,2,\cdots,m_{r}$, the following equality is valid
\begin{eqnarray}
& &\hspace*{-1cm} G_{yv}(\lambda_{r,i}) \!+\! \widehat{\overline{H}}_{r,i}(\theta)G_{zv}(\lambda_{r,i}) \nonumber \\
&& \hspace*{-1.4cm} = G_{yv}(\lambda_{r,i})\left[ I_{m_{v}} \!-\! P(\theta)G_{zv}(\lambda_{r,i})\right]^{-1}
\!+\! W_{r,i}G_{zv}(\lambda_{r,i})
\label{app-28}
\end{eqnarray}
while for each $i=1,2,\cdots,m_{c}$, the next two relations can be established
\begin{eqnarray}
& & \hspace*{-1.0cm} G_{yv}^{[r]}(\lambda_{c,i}) \!+\! \widehat{\overline{H}}^{[r]}_{c,i}(\theta)G_{zv}^{[r]}(\lambda_{c,i}) \!-\!
\widehat{\overline{H}}^{[i]}_{c,i}(\theta)G_{zv}^{[i]}(\lambda_{c,i})
 \nonumber \\
& & \hspace*{-1.4cm} = G_{yv}^{[r]}(\lambda_{c,i}) {\it R_{e}}\!\!\left\lbrace \left[ I_{m_{v}} \!-\! P(\theta)G_{zv}(\lambda_{c,i})\right]^{-1}\right\rbrace \!-\!  \nonumber \\
& & \hspace*{-1.0cm}
G_{yv}^{[i]}(\lambda_{c,i}) {\it I_{m}}\!\!\left\lbrace \left[ I_{m_{v}} \!-\! P(\theta)G_{zv}(\lambda_{c,i})\right]^{-1}\right\rbrace
\!+\! \nonumber \\
& & \hspace*{-1.0cm} W^{[r]}_{c,i}G^{[r]}_{zv}(\lambda_{c,i}) \!-\! W^{[i]}_{c,i}G^{[i]}_{zv}(\lambda_{c,i})
\label{app-29}
\end{eqnarray}
and
\begin{eqnarray}
& & \hspace*{-1.0cm} G_{yv}^{[i]}(\lambda_{c,i}) \!+\! \widehat{\overline{H}}^{[r]}_{c,i}(\theta)G_{zv}^{[i]}(\lambda_{c,i}) \!+\!
\widehat{\overline{H}}^{[i]}_{c,i}(\theta)G_{zv}^{[r]}(\lambda_{c,i})
 \nonumber \\
& & \hspace*{-1.4cm} = G_{yv}^{[r]}(\lambda_{c,i}) {\it I_{m}}\!\!\left\lbrace \left[ I_{m_{v}} \!-\! P(\theta)G_{zv}(\lambda_{c,i})\right]^{-1}\right\rbrace \!+\!  \nonumber \\
& & \hspace*{-1.0cm}
G_{yv}^{[i]}(\lambda_{c,i}) {\it R_{e}}\!\!\left\lbrace \left[ I_{m_{v}} \!-\! P(\theta)G_{zv}(\lambda_{c,i})\right]^{-1}\right\rbrace
\!+\! \nonumber \\
& & \hspace*{-1.0cm} W^{[r]}_{c,i}G^{[i]}_{zv}(\lambda_{c,i}) \!+\! W^{[i]}_{c,i}G^{[r]}_{zv}(\lambda_{c,i})
\label{app-30}
\end{eqnarray}

Define matrices $\Psi_{g}$, $\Psi_{p}$ and $\Psi_{w}$ respectively as follows
\begin{eqnarray*}
& & \hspace*{-0.5cm} \Psi_{p} \!\!=\!\! \left[ {\it vec}(P_{1})\;\; {\it vec}(P_{2}) \;\; \cdots \;\; {\it vec}(P_{m_{\theta}})\right]  \nonumber \\
& & \hspace*{-0.5cm} \Psi_{w} \!\!=\!\! \left[\!\!\!\! \begin{array}{c}
{\it col}\!\!\left\{\!\!\left.\left[I\otimes \left( W_{r,i}G_{zv}(\lambda_{r,i}) \right)\right]\right|_{i=1}^{m_{r}} \!\!\right\} \\
{\it col}\!\!\left\{\!\!\left.\left[\!\!\!\! \begin{array}{l}
I\otimes \left( W^{[r]}_{c,i}G^{[r]}_{zv}(\lambda_{c,i}) \!-\! W^{[i]}_{c,i}G^{[i]}_{zv}(\lambda_{c,i}) \right) \\
I\otimes \left( W^{[r]}_{c,i}G^{[i]}_{zv}(\lambda_{c,i}) \!+\! W^{[i]}_{c,i}G^{[r]}_{zv}(\lambda_{c,i}) \right)
\end{array} \!\!\!\!\right]\!\!\right|_{i=1}^{m_{c}} \right\} \end{array}\!\!\!\!\right]   \nonumber \\
& & \hspace*{-0.5cm} \Psi_{g} \!\!=\!\! \left[\!\!\!\! \begin{array}{c}
{\it col}\!\!\left\{\!\left.\left[I\otimes \left( G_{yv}(\lambda_{r,i})\left[ I_{m_{v}} \!\!-\! P(\theta)G_{zv}(\lambda_{r,i})\right]^{-1} \right)\right]\right|_{i=1}^{m_{r}} \!\right\} \\
\\
{\it col}\!\!\left\{\!\!\!\!\left.\left[\!\!\!\! \begin{array}{l}
I \!\otimes\!\! \left(\!\! {\it R_{e}}\!\!\left\lbrace\! G_{yv}(\lambda_{c,i})\!\!\left[ I_{m_{v}} \!\!-\! P(\theta)G_{zv}(\lambda_{c,i})\right]^{\!-\!1}\!\right\rbrace \!\right) \\
I \!\otimes\!\! \left(\!\! {\it I_{m}}\!\!\left\lbrace\! G_{yv}(\lambda_{c,i})\!\!\left[ I_{m_{v}} \!\!-\! P(\theta)G_{zv}(\lambda_{c,i})\right]^{\!-\!1}\!\right\rbrace \!\right)
\end{array} \!\!\!\!\!\right]\!\!\right|_{i=1}^{m_{c}} \!\!\right\} \end{array}\!\!\!\!\!\right]
\end{eqnarray*}

Note that
\begin{eqnarray*}
& & \hspace*{-1.0cm} {\it R_{e}}\!\!\left\lbrace\! G_{yv}(\lambda_{c,i})\!\!\left[ I_{m_{v}} \!-\! P(\theta)G_{zv}(\lambda_{c,i})\right]^{\!-\!1}\!\right\rbrace
 \nonumber \\
& & \hspace*{-1.4cm} = G_{yv}^{[r]}(\lambda_{c,i}) {\it R_{e}}\!\!\left\lbrace \left[ I_{m_{v}} \!-\! P(\theta)G_{zv}(\lambda_{c,i})\right]^{-1}\right\rbrace \!-\!  \nonumber \\
& & \hspace*{-1.0cm}
G_{yv}^{[i]}(\lambda_{c,i}) {\it I_{m}}\!\!\left\lbrace \left[ I_{m_{v}} \!-\! P(\theta)G_{zv}(\lambda_{c,i})\right]^{-1}\right\rbrace
\end{eqnarray*}
and
\begin{eqnarray*}
& & \hspace*{-1.0cm} {\it I_{m}}\!\!\left\lbrace\! G_{yv}(\lambda_{c,i})\!\!\left[ I_{m_{v}} \!-\! P(\theta)G_{zv}(\lambda_{c,i})\right]^{\!-\!1}\!\right\rbrace
 \nonumber \\
& & \hspace*{-1.4cm} = G_{yv}^{[r]}(\lambda_{c,i}) {\it I_{m}}\!\!\left\lbrace \left[ I_{m_{v}} \!-\! P(\theta)G_{zv}(\lambda_{c,i})\right]^{-1}\right\rbrace \!+\!  \nonumber \\
& & \hspace*{-1.0cm}
G_{yv}^{[i]}(\lambda_{c,i}) {\it R_{e}}\!\!\left\lbrace \left[ I_{m_{v}} \!-\! P(\theta)G_{zv}(\lambda_{c,i})\right]^{-1}\right\rbrace
\end{eqnarray*}
Then on the basis of Equations (\ref{app-28})-(\ref{app-30}), as well as the definition of the matrix $\Psi$ given immediately after Equation (\ref{eqn:par-est-4}), it can be directly proven that
\begin{equation}
\Psi = \left[\Psi_{g} + \Psi_{w} \right] \Psi_{p}
\label{app-31}
\end{equation}

Similar to the arguments of Equation (\ref{app-24}), it can be proved that to guarantee that the matrix $\Psi$ is of FCR, it is necessary that the matrix $\Psi_{p}$ is of FCR. In addition, it can be declared from Lemma \ref{lemma:1} that when this condition is satisfied, the matrix $\Psi$ is of FCR, if and only if the matrix ${\it col}\!\{ \Psi_{g} + \Psi_{w}, \; \Psi_{p,l}^{\perp} \}$ is of FCR. Moreover, the matrix $\Psi_{g} \Psi_{p}$ is of FCR, if and only if the matrix ${\it col}\!\{ \Psi_{g}, \; \Psi_{p,l}^{\perp} \}$is of FCR.

Note that
\begin{equation}
\left[\begin{array}{c} \Psi_{g} + \Psi_{w} \\
\Psi_{p,l}^{\perp} \end{array}
 \right] = \left[\begin{array}{c} \Psi_{g}  \\
\Psi_{p,l}^{\perp} \end{array}
 \right] + \left[\begin{array}{c}  \Psi_{w} \\
0 \end{array}
 \right]
\label{app-32}
\end{equation}
Recall that a real matrix is of FCR, if and only if the product of its transpose and itself is positive definite. In addition, each eigenvalue of a square matrix is a continuous function of its elements \cite{gv1989,hj2013}. It can therefore be declared from Equation (\ref{app-32}) and the definition of the matrix $\Psi_{w}$ that, if the matrix $\Psi_{g} \Psi_{p}$ is of FCR, then there exists a $\varepsilon >0$, such that with each $||W_{r,i}||\leq \varepsilon$, $i=1,2,\cdots,m_{r}$, and each $||W^{[r]}_{c,i} + j W^{[i]}_{c,i}||\leq \varepsilon$, $i=1,2,\cdots,m_{c}$, the associated matrix $\Psi$ is also of FCR, in which $||\cdot||$ stands for an arbitrary matrix norm. On the contrary, if the matrix $\Psi$ is of FCR, then when both $||W_{r,i}||$ with $i=1,2,\cdots,m_{r}$, and $||W^{[r]}_{c,i} + j W^{[i]}_{c,i}||$ with $i=1,2,\cdots,m_{c}$, are sufficiently small, the matrix $\Psi_{g} \Psi_{p}$ is certainly also of FCR.

On the other hand, similar to Lemma 6 of \cite{zly2024}, it can be proved that the descriptor system $\mathbf{\Sigma}_{p}$ is identifiable with its TFM value at $\lambda_{\star,i}$ with $i=1,2,\cdots, m_{\star}$ and $\star = r, c$, if and only if for each vector pairs $\theta,\, \widetilde{\theta} \in \Theta$, $H(\lambda_{\star,i},\theta) = H(\lambda_{\star,i},\widetilde{\theta})$ for each $i=1,2,\cdots, m_{\star}$ and $\star = r, c$, implies that $\theta = \widetilde{\theta}$.

From Equation (\ref{decom-5}), it is clear that $H(\lambda_{\star,i},\theta) = H(\lambda_{\star,i},\widetilde{\theta})$, if and only if
\begin{eqnarray}
& & \hspace*{-1.5cm}G_{yv}(\lambda_{\star,i})\!\!\left[ I_{m_{v}}-P(\theta)G_{zv}(\lambda_{\star,i})\right]^{-1}\left(P(\widetilde{\theta})-P(\theta) \right)\times \nonumber\\
& & \hspace*{0.6cm}
\left[ I_{m_{z}}-G_{zv}(\lambda_{\star,i})P(\widetilde{\theta})\right]^{-1}
\!\! G_{zu}(\lambda_{\star,i})=0
\label{app-33}
\end{eqnarray}

Moreover, on the basis of its block diagonal structure, straightforward algebraic arguments show that the matrix $G_{zu}(\mathbf{\Sigma}_{s})$ is of FRR, if and only if for each $i=1,2,\cdots, m_{\star}$ and $\star = r, c$, the matrix $G_{zu}(\lambda_{\star,i})$ is of FRR. In addition, investigations in \cite{zy2022} and \cite{zly2024} have clarified that Assumptions \ref{assum:1} and \ref{assum:2} guarantee the invertibility of the matrix $I_{m_{z}} \!\!-\!G_{zv}(s)P(\widetilde{\theta})$. These mean that when the matrix $G_{zu}(\mathbf{\Sigma}_{s})$ is of FRR, the matrix $[ I_{m_{z}} \!\!-\! G_{zv}(\lambda_{\star,i})P(\widetilde{\theta})]^{\!-\!1}
\! G_{zu}(\lambda_{\star,i})$ is also of FRR. Hence, using the parametrization of Equation (\ref{plant-4}), Equation (\ref{app-33}) can be proved to be equivalent to
\begin{equation}
\hspace*{-0.25cm}G_{yv}(\lambda_{\star,i})\!\!\left[ I_{m_{v}}-P(\theta)G_{zv}(\lambda_{\star,i})\right]^{-1}\sum_{i=1}^{m_{\theta}}(\widetilde{\theta}_{k}-\theta_{k})P_{k} =0
\label{app-34}
\end{equation}

Vectorize both sides of the above equation. It can be directly shown from the definitions of the matrices $\Psi_{g}$ and $\Psi_{p}$ that, Equation (\ref{app-34}) is valid for each $i=1,2,\cdots, m_{\star}$ and $\star = r, c$, if and only if
\begin{equation}
\Psi_{g}\Psi_{p}(\widetilde{\theta}-\theta) = 0
\label{app-35}
\end{equation}
Therefore, the descriptor system $\mathbf{\Sigma}_{p}$ is identifiable with its TFM value at $\lambda_{\star,i}$ with $i=1,2,\cdots, m_{\star}$ and $\star = r, c$, if and only if the matrix $\Psi_{g}\Psi_{p}$ is of FCR. This completes the proof. \hspace{\fill}$\Diamond$

\hspace*{-0.40cm}{\rm\bf Proof of Theorem \ref{theo:3}.} From the assumption that $G_{zu}(\mathbf{\Sigma}_{s})$ is of FRR, the assumption of Equation (\ref{eqn:pro-1}) and the equality given by Equation (\ref{app-23}), we have that for each $N \geq N_{\overline{h}}$,
\begin{eqnarray}
& & \overline{U}(t_{1:N}) \overline{U}^{T}(t_{1:N}) \nonumber\\
&=& G_{zu}(\mathbf{\Sigma}_{s})\left(\widetilde{U}(t_{1:N})\widetilde{U}^{T}(t_{1:N})\right) G^{T}_{zu}(\mathbf{\Sigma}_{s}) \nonumber\\
&\geq & G_{zu}(\mathbf{\Sigma}_{s})\left(f_{s}(N) I_{m_{u}}\right) G^{T}_{zu}(\mathbf{\Sigma}_{s})
\nonumber\\
&\geq & \underline{\sigma}^{2}(G_{zu}(\mathbf{\Sigma}_{s})) f_{s}(N) I_{m_{z}m_{\xi}}
\label{app-36}
\end{eqnarray}

As both $\underline{\sigma}(G_{zu}(\mathbf{\Sigma}_{s}))$ and $f_{s}(N)$ are positive, the last inequality in the above equation means that $\overline{U}(t_{1:N})$ is of FRR, and
\begin{equation}
\underline{\sigma}(\overline{U}(t_{1:N})) \geq \underline{\sigma}(G_{zu}(\mathbf{\Sigma}_{s})) \sqrt{f_{s}(N)}
\label{app-37-a}
\end{equation}
In addition, from the definition of the Moore-Penrose inverse of a matrix \cite{gv1989,hj2013}, it is immediate that
\begin{equation}
\overline{U}^{\dagger}(t_{1:N}) = \overline{U}^{T}(t_{1:N}) \left( \overline{U}(t_{1:N}) \overline{U}^{T}(t_{1:N}) \right)^{-1}
\label{app-37}
\end{equation}

On the basis of this equation, as well as Equations (\ref{eqn:est-3}) and (\ref{eqn:est-3-a}), we further have that
\begin{equation}
\hspace*{-0.25cm} \widehat{\overline{H}}(\theta) \!=\! \overline{H}(\theta) \!+\!
\left[{Y}_{t}(t_{1:N}) \!+\! N(t_{1:N}) \right] \overline{U}^{\dagger}\!(t_{1:N})
\label{app-38}
\end{equation}
Hence,
\begin{eqnarray}
e_{\overline{h}}(N) \!\!\!\! &=& \!\!\!\! {\it vec}\left(\widehat{\overline{H}}(\theta) - \overline{H}(\theta)\right) \nonumber \\
&=& \!\!\!\!\left(\overline{U}^{\dagger,T}(t_{1:N})\otimes I\right)\!\!\left[ y_{t}(t_{1:N})
\!+\! n(t_{1:N}) \right]
\label{app-39}
\end{eqnarray}
in which
\begin{displaymath}
y_{t}(t_{1:N})={\it col}\!\left\{{y}_{t}(t_{k})|_{k=1}^{N}\right\}, \hspace{0.25cm}
n(t_{1:N})={\it col}\!\left\{n(t_{k})|_{k=1}^{N}\right\}
\end{displaymath}
It can therefore be declared from Assumption \ref{assum:5} that
\begin{equation}
\hspace*{-0.25cm}
{\it E_{x}}\!\!\left\{e_{\overline{h}}(N)\right\}= \left(\overline{U}^{\dagger,T}(t_{1:N})\otimes I\right)y_{t}(t_{1:N})
\label{app-40}
\end{equation}

Based on properties of matrix Kronecker products \cite{hj2013,zdg1996}, it can be further shown from Equation (\ref{app-40}) that
\begin{eqnarray}
& & \hspace*{-1.0cm}\left|\left|{\it E_{x}}\!\!\left\{e_{\overline{h}}(N)\right\}\right|\right|_{2}^{2} \nonumber \\
& & \hspace*{-1.4cm}
= y^{T}_{t}(t_{1:N})
\left[\left(\overline{U}^{\dagger}(t_{1:N})\overline{U}^{\dagger,T}(t_{1:N})\right)\otimes I\right]
y_{t}(t_{1:N})
\label{app-41}
\end{eqnarray}

Recall that the matrix $\overline{U}(t_{1:N})$ has been proved to be FRR. From matrix theories \cite{hj2013} and Equation (\ref{app-36}), it can be declared that there exist orthogonal matrices $U_{\overline{u}}$ and $V_{\overline{u}}=[V_{\overline{u},1}\;\;V_{\overline{u},2}]$, as well as a diagonal matrix $\Sigma_{\overline{u}}$, such that
\begin{equation}
\overline{U}(t_{1:N})= U_{\overline{u}}\Sigma_{\overline{u}}V^{T}_{\overline{u},1},\hspace{0.25cm} \Sigma_{\overline{u}} \geq \underline{\sigma}(G_{zu}(\mathbf{\Sigma}_{s})) \sqrt{f_{s}(N)} I_{m_{z}m_{\xi}}
\label{app-42}
\end{equation}

On the basis of this expression and Equation (\ref{app-37}), the following equality is established.
\begin{equation}
\overline{U}^{\dagger}(t_{1:N})\overline{U}^{\dagger,T}(t_{1:N}) =
V_{\overline{u},1} \Sigma^{-2}_{\overline{u}} V^{T}_{\overline{u},1}
\label{app-43}
\end{equation}
Note that $V_{\overline{u},1}V^{T}_{\overline{u},1} = I -V_{\overline{u},2}V^{T}_{\overline{u},2} \leq I$ and $0 < \Sigma^{-2}_{\overline{u}} \leq \underline{\sigma}^{-2}(\overline{U}(t_{1:N}))I$. We therefore have that
\begin{eqnarray}
\overline{U}^{\dagger}(t_{1:N})\overline{U}^{\dagger,T}(t_{1:N}) & \leq &
V_{\overline{u},1} \!\!
\left(\underline{\sigma}^{-2}(\overline{U}(t_{1:N})) I \!\right) \!\!
V^{T}_{\overline{u},1}
\nonumber \\
& \leq & \underline{\sigma}^{-2}(\overline{U}(t_{1:N}) I
\label{app-44}
\end{eqnarray}

Moreover, from Equation (\ref{eqn:pro-1}), standard results in matrix analysis \cite{hj2013} lead to the following inequality,
\begin{equation}
\underline{\sigma}(\widetilde{U}(t_{1:N})) \geq \sqrt{f_{s}(N)}
\label{app-46}
\end{equation}
Furthermore, it is immediate from Equation (\ref{app-36}) that
\begin{equation}
\underline{\sigma}(\overline{U}(t_{1:N})) \geq \underline{\sigma}(G_{zu}(\mathbf{\Sigma}_{s})) \underline{\sigma}(\widetilde{U}(t_{1:N}))
\label{app-47}
\end{equation}

Substitute Equations (\ref{app-46}) and (\ref{app-47}) into Equation (\ref{app-44}), and define a positive number $\kappa_{zu}$ as
$\kappa_{zu} = \underline{\sigma}^{-2}(G_{zu}(\mathbf{\Sigma}_{s}))$. Then the following inequalities are obtained immediately
\begin{eqnarray}
\overline{U}^{\dagger}(t_{1:N})\overline{U}^{\dagger,T}(t_{1:N}) \!\!\!\!
& \leq & \!\!\!\! \frac{1}
{\underline{\sigma}^{2}(G_{zu}(\mathbf{\Sigma}_{s})) \underline{\sigma}^{2}(\widetilde{U}(t_{1:N}))}I \nonumber\\
& \leq & \!\!\!\!
\frac{\kappa_{zu}}{f_{s}(N)}I
\label{app-48}
\end{eqnarray}

Note that the matrix $G_{zu}(\mathbf{\Sigma}_{s})$ is assumed to be FRR, and its dimension does not depend on the data length $N$. It is clear that $\kappa_{zu}$ is a finite and $N$ independent number.

On the other hand, from the assumption that the descriptor system $\mathbf{\Sigma}_{p}$ is stable, as well as the expression for its transient response $y_{t}(t)$ that is given by Equation (\ref{decom-2-1}), it is clear that there exist two positive constants $\kappa_{t}$ and $\lambda_{t}$ that do not depend on the data length $N$, such that for each $k=1,2,\cdots,N$,
\begin{equation}
\left|\left|y_{t}(t_{k})\right|\right|_{2} \leq \kappa_{t}e^{-\lambda_{t} t_{k}}
\label{app-49}
\end{equation}

Define $t_{d}$ as
\begin{displaymath}
t_{d} = \min_{2 \leq k \leq N}\left( t_{k} - t_{k-1}\right)
\end{displaymath}
that is, the minimal interval between two succeeding sampling instants. It is obvious that $t_{d} > 0$. Based on Equations (\ref{app-41}) and (\ref{app-49}), as well as the last inequality of Equation (\ref{app-48}), we have that
\begin{eqnarray}
\hspace*{-1.0cm}\left|\left|{\it E_{x}}\!\!\left\{e_{\overline{h}}(N)\right\}\right|\right|_{2}^{2} \!\! &\leq& \!\! y^{T}_{t}(t_{1:N})
\left(\frac{\kappa_{zu}}{f_{s}(N)} I\right)
y_{t}(t_{1:N}) \nonumber \\
&=& \!\!
\frac{\kappa_{zu}}{f_{s}(N)}\sum_{k=1}^{N} y^{T}_{t}(t_{k})
y_{t}(t_{k}) \nonumber \\
&\leq& \!\! \frac{\kappa_{zu}}{f_{s}(N)}\sum_{k=1}^{N} \kappa^{2}_{t}e^{-2\lambda_{t} t_{k}} \nonumber \\
&\leq& \!\! \frac{\kappa_{zu} \kappa^{2}_{t} e^{-2\lambda_{t} t_{1}}}{f_{s}(N)}\sum_{k=0}^{N-1} e^{-2\lambda_{t} t_{d}k} \nonumber \\
&\leq& \!\! \frac{\kappa_{zu}\kappa^{2}_{t}e^{-2\lambda_{t} t_{1}}}{1-e^{-2\lambda_{t} t_{d}}} \times \frac{1}{f_{s}(N)}
\label{app-50}
\end{eqnarray}

Define $\kappa_{\overline{h},e}$ as
\begin{displaymath}
\kappa_{\overline{h},e} = \sqrt{\frac{\kappa_{zu}e^{-2\lambda_{t} t_{1}}}{1-e^{-2\lambda_{t} t_{d}}}}\kappa_{t}
\end{displaymath}
Then it is clear from the properties of the associated numbers that, $\kappa_{\overline{h},e}$ is independent of the data length $N$. In addition, the last inequality of Equation (\ref{app-50}) implies that it satisfies Equation (\ref{eqn:theo5-1}).

Now, we prove the second property of the estimate $\widehat{\overline{H}}(\theta)$. From Equations (\ref{app-39}) and (\ref{app-40}), it is immediate that
\begin{equation}
e_{\overline{h}}(N) \!-\! {\it E_{x}}\!\!\left\{e_{\overline{h}}(N)\right\} \!\! =\!\! \left(\overline{U}^{\dagger,T}(t_{1:N})\otimes I\right)\! n(t_{1:N})
\label{app-51}
\end{equation}

We therefore from Assumption \ref{assum:5} have that
\begin{eqnarray}
& & \hspace*{-0.6cm} {\it E_{x}}\!\!\left\{\!\!\left[e_{\overline{h}}(N) \!-\!\! {\it E_{x}}\!\!\left(\!e_{\overline{h}}(N)\!\right)\!\right] \!\!\!\left[ e_{\overline{h}}(N) \!-\!\! {\it E_{x}}\!\!\left(\!e_{\overline{h}}(N) \!\right)\! \right]^{\!T}\right\} \nonumber \\
& & \hspace*{-1.0cm}=\! \left(\!\overline{U}^{\dagger,T}(t_{1:N}) \!\otimes\! I \!\right)\!\!
{\it E_{x}}\!\!\left\{\! n(t_{1:N}) n^{T}\!(t_{1:N}) \!\right\}\!
\!\left(\!\overline{U}^{\dagger,T}(t_{1:N}) \!\otimes\! I \!\right)^{T}  \nonumber \\
& & \hspace*{-1.0cm}\leq\! \left(\overline{U}^{\dagger,T}(t_{1:N})\otimes I\right)\!
\! \left( I \otimes \Sigma_{n}\right)
\left(\overline{U}^{\dagger,T}(t_{1:N})\otimes I\right)^{T} \nonumber \\
& & \hspace*{-1.0cm}=\! \left(\overline{U}^{\dagger,T}(t_{1:N})\overline{U}^{\dagger}(t_{1:N})\right)\otimes \Sigma_{n}
\label{app-52}
\end{eqnarray}

On the other hand, from Equation (\ref{app-48}) and the definition of the matrix $\Sigma_{n}$, we have that
\begin{eqnarray}
\left(\!\overline{U}^{\dagger,T}\!\!(t_{1:N})\overline{U}^{\dagger}\!(t_{1:N})\!\right) \!\otimes\! \Sigma_{n} \!\!\!\!&\leq&\!\!\!\!
\left(\!\overline{U}^{\dagger,T}\!\!(t_{1:N}) \overline{U}^{\dagger}\!(t_{1:N})\!\right) \!\otimes\! \left( \overline{\sigma}(\Sigma_{n})I \right) \nonumber  \\
&\leq&\!\!\!\! \frac{\kappa_{zu}\overline{\sigma}(\Sigma_{n})}{f_{s}(N)}I
\label{app-53}
\end{eqnarray}
A combination of Equations (\ref{app-52}) and (\ref{app-53}) implies that $\kappa_{\overline{h},c}=\kappa_{zu}$ satisfies the requirements of Equation (\ref{eqn:theo5-2}).

Now, we prove the third property of the estimate $\widehat{\overline{H}}(\theta)$. From Equation (\ref{app-39}) and Assumption \ref{assum:5}, we have that
\begin{eqnarray}
\hspace*{-0.0cm} {\it E_{x}}\!\!\left\{\!\!e_{\overline{h}}(N) e^{T}_{\overline{h}}(N) \!\!\right\} \!\!\!\!\!\!&=&\!\!\!\!\!\! \left(\!\overline{U}^{\dagger,T}(t_{1:N}) \!\otimes\! I \!\right)\!\!
{\it E_{x}}\!\!\left\{\! y(t_{1:N}) y^{T}\!(t_{1:N}) + \right. \nonumber \\
& & \hspace*{0.0cm}
\! y(t_{1:N}) n^{T}\!(t_{1:N}) \!+\! n(t_{1:N}) y^{T}\!(t_{1:N}) +\nonumber \\
& & \hspace*{0.35cm}\left. n(t_{1:N}) n^{T}\!(t_{1:N}) \!\right\}\!
\!\left(\!\overline{U}^{\dagger,T}(t_{1:N}) \!\otimes\! I \!\right)^{T}  \nonumber \\
&\leq& \!\!\!\!\!\! \left(\overline{U}^{\dagger,T}(t_{1:N})\otimes I\right)\!
\! \left( y(t_{1:N}) y^{T}\!(t_{1:N}) + \right. \nonumber \\
& & \hspace*{0.35cm} \left. I \otimes \Sigma_{n}\right)
\left(\overline{U}^{\dagger}(t_{1:N})\otimes I\right)
\label{app-54}
\end{eqnarray}
Hence
\begin{eqnarray}
\hspace*{-0.0cm} {\it E_{x}}\!\!\left\{\!e^{T}_{\overline{h}}(N) e_{\overline{h}}(N) \!\right\} \!\!\!\!\!\!&=&\!\!\!\!\!\!
{\it E_{x}}\!\!\left\{\!\!{\it tr}\!\!\left(e_{\overline{h}}(N) e^{T}_{\overline{h}}(N)\right) \!\!\right\}
\nonumber \\
&=&\!\!\!\!\!\!
{\it tr}\!\!\left({\it E_{x}}\!\!\left\{\!e_{\overline{h}}(N) e^{T}_{\overline{h}}(N) \!\!\right\} \right) \nonumber \\
&\leq& \!\!\!\!\!\!
y^{T}_{t}\!(t_{1:N})\!\!
\left[\!\!\left(\!\overline{U}^{\dagger}(t_{1:N})\overline{U}^{\dagger,T}\!(t_{1:N})\!\!\right)\!\!\otimes\! I \!\right]\!y_{t}(t_{1:N}) \!+ \nonumber \\
& & \hspace*{-0.30cm}
{\it tr}\!\!\left\{\!\!\left(\overline{U}^{\dagger,T}\!\!(t_{1:N})\overline{U}^{\dagger}(t_{1:N})\right) \!\otimes\! \Sigma_{n} \!\right\}
\label{app-55}
\end{eqnarray}

Substitute Equations (\ref{app-41}), (\ref{app-50}) and (\ref{app-53}) into the above equation, and take the dimensions of the associated identity matrices into account. The following inequality is straightforwardly obtained,
\begin{eqnarray}
\hspace*{-0.05cm} {\it E_{x}}\!\!\left\{\!e^{T}_{\overline{h}}(N) e_{\overline{h}}(N) \!\right\} \!\!\!\!\!\!
&\leq& \!\!\!\!\!\!
\frac{\kappa_{zu} \kappa^{2}_{t} e^{-2\lambda_{t} t_{1}}}{(1 \!-\! e^{-2\lambda_{t} t_{d}})f_{s}(N)}
 \!+\! \frac{\kappa_{zu} \overline{\sigma}(\Sigma_{n})m_{y}m_{z}m_{\xi}}{f_{s}(N)}\nonumber \\
&=& \!\!\!\!\!\!
\frac{\kappa_{\overline{h},c}\left[m_{y}m_{z}m_{\xi}\overline{\sigma}(\Sigma_{n}) \!+\! \frac{\kappa^{2}_{t}e^{-2\lambda_{t} t_{1}}}{1-e^{-2\lambda_{t} t_{d}}}\right]}{f_{s}(N)}
\label{app-56}
\end{eqnarray}
The proof can now be completed by letting $\kappa_{\overline{h},m} = \frac{\kappa^{2}_{t}e^{-2\lambda_{t} t_{1}}}{1-e^{-2\lambda_{t} t_{d}}}$.
\hspace{\fill}$\Diamond$

\hspace*{-0.40cm}{\rm\bf Proof of Theorem \ref{theo:4}.} For brevity, define a vector $w(N)$ as
\begin{displaymath}
w(N) = {\it col}\!\!\left\{\!\! \left.{\it vec}(W_{r,i})\right|_{i=1}^{m_{r}}, \left.
\left[\!\!\begin{array}{c} {\it vec}(W^{[r]}_{c,i})  \\ {\it vec}(W^{[i]}_{c,i}) \end{array}\!\!\right]\!\right|_{i=1}^{m_{c}} \!\!\right\}
\end{displaymath}
Moreover, denote $W^{[r]}_{c,i}+jW^{[i]}_{c,i}$ by $W_{c,i}$ for each $i=1,2,\cdots,m_{c}$. Then from Equations (\ref{eqn:est-1}) and (\ref{eqn:est-2}), it is immediate that for every $i=1,2,\cdots,m_{\star}$ with $\star=r,c$, the following equality is valid,
\begin{equation}
\widehat{\overline{H}}_{\star,i}(\theta) \!=\! G_{yv}(\lambda_{\star,i})\left[ I_{m_{v}} \!-\! P(\theta)G_{zv}(\lambda_{\star,i})\right]^{-1} P(\theta) \!-\! W_{\star,i}
\label{app-57}
\end{equation}

On the basis of this equality and the definition of the vector $\overline{h}$ given immediately after Equation (\ref{eqn:par-est-4}), as well as the definitions of the matrices $\Psi_{g}$ and $\Psi_{p}$ given immediately after Equation (\ref{app-30}), some straightforward algebraic manipulations show that
\begin{equation}
\overline{h} = \Psi_{g}\Psi_{p} \theta - w(N)
\label{app-58}
\end{equation}

Substitute this equation into Equation (\ref{eqn:par-est-5}), and recall from Equation (\ref{app-31}) that $\Psi =(\Psi_{g} + \Psi_{w})\Psi_{p}$. The following relation is obtained,
\begin{eqnarray}
\widehat{\theta} &=& (\Psi^{T}\Psi)^{-1} \Psi^{T}\left[(\Psi - \Psi_{w} \Psi_{p}) \theta -w(N) \right] \nonumber\\
&=& \theta - (\Psi^{T}\Psi)^{-1} \Psi^{T}\left[\Psi_{w} \Psi_{p} \theta + w(N) \right]
\label{app-59}
\end{eqnarray}

On the other hand, note that ${\it vec}\{P(\theta)\} = \Psi_{p}\theta$. It can be directly shown that
\begin{eqnarray*}
& & \hspace*{-0.5cm} \left[I_{m_{z}} \!\otimes\! \left( W_{r,i}G_{zv}(\lambda_{r,i}) \right)\right] \!\Psi_{p}\theta \nonumber \\
& & \hspace*{0.0cm}=\!\!
{\it vec}\{W_{r,i}G_{zv}(\lambda_{r,i})P(\theta)\} \nonumber \\
& & \hspace*{0.0cm}=\!\! \left(P^{T}(\theta)\!\otimes\! I_{m_{y}} \!\right)\!\! \left(G^{T}_{zv}(\lambda_{r,i}) \!\otimes\! I_{m_{y}} \!\right){\it vec}\{W_{r,i}\} \nonumber\\
& & \hspace*{-0.5cm} \left[I_{m_{z}} \!\otimes\! \left( W^{[r]}_{c,i}G^{[r]}_{zv}(\lambda_{c,i}) \!-\! W^{[i]}_{c,i}G^{[i]}_{zv}(\lambda_{c,i}) \right)\right] \!\Psi_{p}\theta \nonumber \\
& & \hspace*{0.0cm}=\!\!
{\it vec}\left\{\left(W^{[r]}_{c,i}G^{[r]}_{zv}(\lambda_{c,i}) \!-\! W^{[i]}_{c,i}G^{[i]}_{zv}(\lambda_{c,i})\right)P(\theta)\right\}
\end{eqnarray*}

\begin{eqnarray*}
& & \hspace*{0.0cm}=\!\! \left(P^{T}(\theta)\!\otimes\! I_{m_{y}} \!\right)\!\!\left[\!G^{[r],T}_{zv}(\lambda_{c,i}) \!\otimes\! I_{m_{y}} \;\;
-\! G^{[i],T}_{zv}(\lambda_{c,i}) \!\otimes\! I_{m_{y}} \!\right] \!\!\times \nonumber \\
& & \hspace*{4.10cm}
{\it col}\!\!\left\{{\it vec}\{W^{[r]}_{c,i}\}, \; {\it vec}\{W^{[i]}_{c,i}\} \right\}\nonumber\\
& & \hspace*{-0.5cm} \left[I_{m_{z}} \!\otimes\! \left( W^{[r]}_{c,i}G^{[i]}_{zv}(\lambda_{c,i}) \!+\! W^{[i]}_{c,i}G^{[r]}_{zv}(\lambda_{c,i}) \right)\right] \!\Psi_{p}\theta \nonumber \\
& & \hspace*{0.0cm}=\!\!
{\it vec}\left\{\left(W^{[r]}_{c,i}G^{[i]}_{zv}(\lambda_{c,i}) \!+\! W^{[i]}_{c,i}G^{[r]}_{zv}(\lambda_{c,i})\right)P(\theta)\right\} \nonumber \\
& & \hspace*{0.0cm}=\!\! \left(P^{T}(\theta)\!\otimes\! I_{m_{y}} \!\right)\!\!\left[\!G^{[i],T}_{zv}(\lambda_{c,i}) \!\otimes\! I_{m_{y}} \;\;\;
 G^{[r],T}_{zv}(\lambda_{c,i}) \!\otimes\! I_{m_{y}} \!\right] \!\!\times \nonumber \\
& & \hspace*{4.1cm}
{\it col}\!\!\left\{{\it vec}\{W^{[r]}_{c,i}\}, \; {\it vec}\{W^{[i]}_{c,i}\} \right\}
\end{eqnarray*}

Similar to the matrix $G_{zu}(\mathbf{\Sigma}_{s})$, define a block diagonal real matrix $G_{zv}(\mathbf{\Sigma}_{s})$ as follows,
\begin{eqnarray*}
& & \hspace*{-1.0cm} G_{zv}(\mathbf{\Sigma}_{s}) =
{\it diag}\!\!\left\{ {\it diag}\!\!\left\{\left. G^{T}_{zv}(\lambda_{r,i})\right|_{i=1}^{m_{r}}\right\}, \right. \nonumber\\
& & \hspace*{1.4cm} \left.
{\it diag}\!\!\!\left\{\!\!\left[\!\!\begin{array}{rr}
G_{zv}^{[r],T}(\lambda_{c,i}) & -G_{zv}^{[i],T}(\lambda_{c,i}) \\
G_{zv}^{[i],T}(\lambda_{c,i}) & G_{zv}^{[r],T}(\lambda_{c,i})\end{array}\right]_{i=1}^{m_{c}}\!\right\}\!\!\right\}
\end{eqnarray*}
Then based on the above three equalities and the definitions of the matrices $\Psi_{w}$ given immediately after Equation (\ref{app-30}), we have that
\begin{equation}
\Psi_{w} \Psi_{p} \theta = \left(I_{m_{\xi}} \!\otimes\! P^{T}(\theta) \!\otimes\! I_{m_{y}}\right) G_{zv}(\mathbf{\Sigma}_{s}) w(N)
\label{app-60}
\end{equation}

From the definition of $\overline{H}(\theta)$ given immediately after Equation (\ref{eqn:out-meas-2}) and that of the vector $e_{\overline{h}}(N)$ given by Equation (\ref{eqn:pro-2}), it is clear that
\begin{equation}
 w(N) = -e_{\overline{h}}(N)
\label{app-62}
\end{equation}
Substitute this relation and Equation (\ref{app-60}) into Equation (\ref{app-59}). The following equality is obtained
\begin{equation}
\widehat{\theta} \!-\! \theta =(\Psi^{T}\Psi)^{-1}\!\Psi^{T} \!\!\left[I \!+\! \left(I_{m_{\xi}} \!\otimes\! P^{T}\!(\theta) \!\otimes\! I_{m_{y}} \!\right) \!G_{zv}(\mathbf{\Sigma}_{s}) \!\right] \!e_{\overline{h}}(N)
\label{app-61}
\end{equation}

Let $w_{i}(N)$ denote the $i$-th row element of the vector $w(N)$, $i=1,2,\cdots,m_{y}m_{z}m_{\xi}$. Then when the assumptions of Theorem \ref{theo:3} are simultaneously satisfied and $N>N_{\overline{h}}$, the following inequalities can be established for an arbitrary positive number $\varepsilon$, on the basis of Lemma \ref{lemma:0} and Equations (\ref{app-62}) and (\ref{eqn:theo5-3}),
\begin{eqnarray}
\hspace*{-1.0cm} {\it P_{r}}\!\left\{|w_{i}(N)| > \varepsilon \right\} \!\!\!\!&\leq&\!\!\!\! \frac{{\it E_{x}}\!\!\left\{w_{i}^{2}(N)  \right\}}{\varepsilon^{2} }
\nonumber \\
&\leq& \!\!\!\!\frac{{\it E_{x}}\!\!\left\{e^{T}_{\overline{h}}(N)e_{\overline{h}}(N)\right\} }{\varepsilon^{2} }
\nonumber \\
&\leq& \!\!\!\!
\frac{\kappa_{\overline{h},c}\left[m_{y}m_{z}m_{\xi}\overline{\sigma}(\Sigma_{n}) + \kappa_{\overline{h},m} \right]}{\varepsilon^{2}f_{s}(N)}
\label{app-63}
\end{eqnarray}

Therefore, when the condition of Equation (\ref{eqn:theo6-0}) is satisfied, we have that
\begin{eqnarray}
\hspace*{-1.0cm} & & \sum_{N=1}^{\infty} {\it P_{r}}\!\left\{|w_{i}(N)| > \varepsilon \right\} \nonumber \\
&=&\!\!\!\!
\sum_{N=1}^{N_{\overline{h}}} {\it P_{r}}\!\left\{|w_{i}(N)| > \varepsilon \right\} +
\sum_{N=N_{\overline{h}}+1}^{\infty} \!\!\!\!\!\!{\it P_{r}}\!\left\{|w_{i}(N)| > \varepsilon \right\} \nonumber \\
&\leq&\!\!\!\! N_{\overline{h}} +
\frac{\kappa_{\overline{h},c}\left[m_{y}m_{z}m_{\xi}\overline{\sigma}(\Sigma_{n}) + \kappa_{\overline{h},m} \right]}{\kappa_{s}\varepsilon^{2}}
\sum_{N=N_{\overline{h}}+1}^{\infty}\!\!\!\!\!\! N^{-(1+\delta_{s})} \nonumber \\
&<& \!\!\!\! \infty
\label{app-64}
\end{eqnarray}
Hence, it can be declared from Lemma \ref{lemma:0} that with the increment of the data length $N$, each element of the vector $w(N)$ converges to $0$ w.p.1. That is, for every positive number $\varepsilon$, there exists a finite positive integer $\overline{N}_{\theta}$, such that for each $N \geq \overline{N}_{\theta}$ and each  $i=1,2,\cdots,m_{y}m_{z}m_{\xi}$, the following inequality is valid,
\begin{equation}
{\it P_{r}}\!\left\{|w_{i}(N)| \leq \varepsilon \right\} \!\geq\! 1-\delta_{\theta}
\label{app-65}
\end{equation}

According to Lemma \ref{lemma:4}, these mean that when the descriptor system $\mathbf{\Sigma}_{p}$ is identifiable for each $\theta\in\mathbf{\Theta}$ using its TFM value at $\lambda_{\star,i}$, in which $i=1,2,\cdots, m_{\star}$ and $\star = r, c$, then for each $N \geq \overline{N}_{\theta}$, the matrix $\Psi$ of Equation
(\ref{eqn:par-est-4}) is of FCR with a probability not smaller than $1-\delta_{\theta}$, provided that $\varepsilon$ is sufficiently small.

On the other hand, from the definition of the matrix $G_{zv}(\mathbf{\Sigma}_{s})$, it is clear that if Assumption \ref{assum:7} is satisfied, then every element of the matrix $G_{zv}(\mathbf{\Sigma}_{s})$ has a finite value. Therefore, its maximum singular value holds also this property, noting that for every matrix with a finite dimension, its maximum singular value is not greater than the square root of the summation of the magnitude square of its elements. In addition, when the parameter set $\mathbf{\Theta}$ is bounded, it can be claimed from Equation
(\ref{plant-4}) that the matrix $P(\theta)$ also has a finite maximum singular value.

Recall that the maximum singular value of the product of two matrices is not greater than the product of their maximum singular values \cite{gv1989,hj2013}. The above arguments show that when $f_{s}(N)$ satisfies Equation (\ref{eqn:theo6-0}) and $N \geq \overline{N}_{\theta}$, the probability is not smaller than $1-\delta_{\theta}$ that the maximum singular value of the matrix $(\Psi^{T}\Psi)^{-1}\!\Psi^{T} \![I \!+\! (I_{m_{\xi}} \!\otimes\! P^{T}\!(\theta) \!\otimes\! I_{m_{y}} \!) G_{zv}(\mathbf{\Sigma}_{s}) \!]$ is finite.

The proof can now be completed by Equation
(\ref{app-61}) and the conclusions of Theorem \ref{theo:3}, letting ${N}_{\theta} = \max\{N_{\overline{h}},\;\overline{N}_{\theta}\}$. \hspace{\fill}$\Diamond$

\section*{Appendix II. The One Jordan Block Case}

From the assumption, all the eigenvalues of the matrix $\Xi$ are obviously real and equal to $\lambda_{r}$, and the dimension of the null space of the matrix $\lambda_{r} I_{m_{\xi}} -\Xi$ is equal to $1$. Then according to matrix theories \cite{hj2013}, there exists a real and invertible matrix $T$, such that
\begin{equation}
\Xi \!=\! T \Lambda T^{-1}\!, \hspace{0.15cm}
\Lambda \!=\!\! \left[\!\!\begin{array}{cccccc}
\lambda_{r} & 1 & 0 & \cdots & 0 & 0 \\
0 & \lambda_{r} & 1 &  \cdots & 0  & 0 \\
\vdots & \vdots &  \ddots & \ddots & \ddots & \vdots \\
0 & 0 & 0 &  \cdots & \lambda_{r}  & 1 \\
0 & 0 & 0 &  \cdots & 0  & \lambda_{r}
\end{array}\!\!\right]
\end{equation}

For each $i=1,2,\cdots, m_{\xi}$, denote the $i$-th column vectors of the matrices $T$, $\Pi T$ and $XT$ respectively by $t_{i}$, $\overline{\pi}_{i}$ and $\overline{x}_{i}$. Then similar arguments as those between Equations (\ref{app-4}) and (\ref{app-8}) show that Equation (\ref{decom-1}) is equivalent to
\begin{eqnarray}
& & \label{app-15}
\hspace*{-1.0cm} \lambda_{r}E\overline{x}_{1} = A\overline{x}_{1} + B \overline{\pi}_{1}	\\
& & \label{app-16}
\hspace*{-1.0cm} \lambda_{r}E\overline{x}_{i} + E\overline{x}_{i-1}= A\overline{x}_{i} + B \overline{\pi}_{i}, \hspace{0.25cm} i=2,3,\cdots,m_{\xi}
\end{eqnarray}

When $\lambda_{r}$  is different from
each generalized eigenvalue of the system matrix pair $(E,\;A)$, the matrix $\lambda_{r}E -A$ is invertible. An explicit expression can therefore be derived for $\overline{x}_{1}$ from Equation (\ref{app-15}), which is given as follows.
\begin{equation}
\label{app-17}
\hspace*{-1.0cm} \overline{x}_{1} = \left(\lambda_{r}E -A \right)^{-1}B \overline{\pi}_{1}	
\end{equation}
Based on this expression and an iterative application of Equation (\ref{app-16}), an explicit expression is also obtained for $\overline{x}_{i}$ with $2 \leq i \leq m_{\xi}$.
\begin{eqnarray}
\hspace*{-1.0cm} \overline{x}_{i} \!\!\!\!&=&\!\!\!\! \left(\lambda_{r}E -A \right)^{-1} \left[ B \overline{\pi}_{i} - E\overline{x}_{i-1} \right] \nonumber \\
&=&\!\!\!\! \cdots
\nonumber \\
&=&\!\!\!\! \sum_{k=0}^{i-1}\left[-\left(\lambda_{r}E -A \right)^{-1}E\right]^{k}\left(\lambda_{r}E -A \right)^{-1}B \overline{\pi}_{i-k}
\label{app-18}
\end{eqnarray}
Clearly, this expression includes the results for $i=1$ as a special case.

On the other hand, from the first equality of Equation (\ref{decom-5}), the following equality can be straightforwardly established for each $k=1,2,\cdots$, through some algebraic manipulations.
\begin{equation}
\label{app-19}
\hspace*{-1.0cm}\frac{d^{k}H(s)}{ds^{k}} \!=\! k!C \left[-\left(sE -A \right)^{-1}E\right]^{k}\!\!\left(sE -A \right)^{-1}\!B
\end{equation}

For brevity, denote the value of $\frac{d^{k}H(s)}{ds^{k}}$ at $s=\lambda_{r}$ by $\frac{d^{k}H(\lambda_{r})}{ds^{k}}$  for each positive integer $k$. Based on these relations and the first two equalities of Equation (\ref{app-10}), direct matrix operations show that
\begin{eqnarray} \label{app-20}
\hspace*{-0.0cm}  CX \!+\! D\Pi
\!\!\!\!\!&=&\!\!\!\!\!\!\!
\left[\! H(\lambda_{r})\overline{\pi}_{1}\;\;\; H(\lambda_{r})\overline{\pi}_{2}\!+\!\frac{dH(\lambda_{r})}{ds}\overline{\pi}_{1}
\;\; \cdots \;\; H(\lambda_{r})\overline{\pi}_{m_{\xi}}\!+
\right. \nonumber \\
& & \hspace*{-0.8cm}\left.
 \frac{dH(\lambda_{r})}{ds}\overline{\pi}_{m_{\xi}-1}\!+\!\cdots\!+\!
\frac{d^{m_{\xi}\!-\!1}\!\!H(\lambda_{r})}{ds^{m_{\xi}\!-\!1}} \!\!\times\!\! \frac{\overline{\pi}_{1}}{(m_{\xi}\!-\!1)!}  \!\right]\!\!T^{-\!1}
\end{eqnarray}

It can therefore be declared from Equation (\ref{decom-2-2}) that in this case, the steady-state response $y_{s}(t)$ of the descriptor system $\mathbf{\Sigma}_{p}$ contains information about both the values of its TFM and its derivatives.

When the system matrix $\Xi$ has several different real/complex eigenvalues and is not diagonalizable, the above algebraic manipulations can be performed independently to each of its individual Jordan blocks, and the same conclusions can be achieved.

\end{document}